\documentclass[usegraphicx,usenatbib,useAMS]{mn2e}
\usepackage{subfigure}

\newcommand{\Cc}{{\cal C}}

\newcommand{\CC}{{\bf C}}

\def\fun#1#2{\lower3.6pt\vbox{\baselineskip0pt\lineskip.9pt
\ialign{$\mathsurround=0pt#1\hfil##\hfil$\crcr#2\crcr\sim\crcr}}}
\def\lap{\mathrel{\mathpalette\fun <}}
\def\gap{\mathrel{\mathpalette\fun >}}
\def\mpl{M_{\rm pl}}
\newcommand{\MUNCH}[1]{\relax}

\begin{document}

\title[CMB polarization and inflation]{Considerations in optimizing CMB polarization experiments to constrain inflationary physics} \author[Verde, Peiris \&
Jimenez]{Licia Verde$^{1}$\thanks{{\tt lverde@physics.upenn.edu}},
Hiranya V. Peiris$^{2}$\thanks{Hubble Fellow, {\tt
hiranya@cfcp.uchicago.edu}} and Raul Jimenez$^{1}$\thanks{{\tt
raulj@physics.upenn.edu}} \\ $^1$Dept. of Physics and Astronomy,
University of Pennsylvania, 209 South 33rd Street, Philadelphia,
PA-19104, USA.\\ $^2$Kavli Institute for Cosmological Physics and
Enrico Fermi Institute, University of Chicago, Chicago IL 60637, USA.}

\maketitle

\begin{abstract}

We quantify the limiting factors in optimizing current-technology
cosmic microwave background (CMB) polarization experiments to learn
about inflationary physics. We consider space-based, balloon-borne and
ground-based experiments. We find that foreground contamination and
residuals from foreground subtraction are ultimately the limiting
factors in detecting a primordial gravity wave signal. For full-sky
space-based experiments, these factors hinder the detection of
tensor-to-scalar ratios of $r < 10^{-3}$ on large scales, while for
ground-based experiments these factors impede the ability to apply
delensing techniques. We consider ground-based/balloon-borne
experiments of currently planned or proposed designs and find that it
is possible for a value of $r=0.01$ to be measured at $\sim
3$-$\sigma$ level.  A small space-based CMB polarization experiment,
with current detector technology and full sky coverage, can detect
$r\sim 1\times 10^{-3}$ at the $\sim 3$-$\sigma$ level, but a markedly
improved knowledge of polarized foregrounds is needed. We advocate
using as wide a frequency coverage as possible in order to carry out
foreground subtraction at the percent level, which is necessary to
measure such a small primordial tensor amplitude.

To produce a clearly detectable ($>$3-$\sigma$) tensor component in a
realistic CMB experiment, inflation must either involve large-field
variations, $\Delta\phi \gap 1$ or multi-field/hybrid models.  Hybrid
models can be easily distinguished from large-field models due to
their blue scalar spectral index. Therefore, an observation of a
tensor/scalar ratio and $n < 1$ in future experiments with the
characteristics considered here may be an indication that inflation is
being driven by some physics in which the inflaton cannot be described
as a fundamental field.
\end{abstract}


\section{INTRODUCTION}
\label{sec:intro}

The inflationary paradigm, proposed and elucidated by
\citet{guth:1981,linde:1982,albrecht/steinhardt:1982,sato:1981,
mukhanov/chibisov:1981,hawking:1982,guth/pi:1982,starobinsky:1982,
bardeen/steinhardt/turner:1983,mukhanov/feldman/brandenberger:1992}
has enjoyed resounding success in the current era of precision
cosmology. Not only has its broad-brush prediction of a flat universe
been successful, but the more detailed predictions of the simplest
inflationary models, of Gaussian, adiabatic, and nearly (but not
exactly) scale-invariant perturbations, are also perfectly consistent
with the data. Observations of the cosmic microwave background (CMB),
in particular by the WMAP satellite (see \citet{bennett/etal:2003} and
references therein), have been instrumental in these observational
tests of inflation \citep{peiris/etal:2003, barger/lee/marfatia:2003,
  leach/liddle:2003, kinney/etal:2004}. Particularly exciting is WMAP's
detection of an anti-correlation between CMB temperature and
polarization fluctuations at angular separations $\theta > 2^\circ$
(corresponding to the $TE$ anti-correlation seen on scales $\ell \sim
100-150$), a distinctive signature of adiabatic fluctuations on
super-horizon scales at the epoch of decoupling
\citep{spergel/zaldarriaga:1997}, confirming a fundamental prediction
of the inflationary paradigm. This is probably the least ambiguous
signature of the acausal physics which characterize inflation, unlike
the large-angle correlations in the CMB temperature which are
contaminated by ``foregrounds'' such as the Integrated Sachs-Wolfe
(ISW) effect.

The detection and measurement of a stochastic gravitational-wave
background would provide the next important key test of inflation. The
CMB polarization anisotropy has the potential to reveal the signature
of primordial gravity waves \citep{zaldarriaga/seljak:1997,
kamionkowski/kosowsky/stebbins:1997}; unlike the CMB temperature
anisotropy, it can discriminate between contributions from density
perturbations (a  scalar quantity) and gravity waves (a tensor
quantity). The polarization anisotropy can be decomposed into two
orthogonal modes: $E$-mode polarization is the curl-free component,
generated by both density and gravity-wave perturbations, and $B$-mode
polarization is the divergence-free mode, which can only, to leading
order, be produced by gravity waves. The primordial $B$-mode signal
appears at large angular scales. On smaller scales, another effect,
which is sub-dominant on large scales, kicks in and eventually
dominates: weak gravitational lensing of the $E$-mode converts it into
$B$-mode polarization \citep{zaldarriaga/seljak:1998}. These
considerations motivate the current observational and experimental
effort, such as on-going ground-based experiments
(e.g. QUaD\footnote{QUaD: {\tt
http://www.stanford.edu/group/quest$_-$telescope/}}
\citep{ChurchQUaD}, CLOVER\footnote{CLOVER: {\tt
http://www.mrao.cam.ac.uk/$\sim$act21/clover.html}},
PolarBeaR\footnote{PolarBeaR: {\tt
http://bolo.berkeley.edu/polarbear/index.html}}, QUIET\footnote{QUIET:\\
{\tt http://cfcp.uchicago.edu/$\sim$peterh/polarimetry/quiet3.html}})
and planned or proposed space-based and balloon-borne missions
(e.g. Planck\footnote{Planck:\\ {\tt
http://sci.esa.int/science-e/www/area/index.cfm?fareaid=17}},
SPIDER\footnote{SPIDER: {\tt http://www.astro.caltech.edu/$\sim$lgg/spider$_-$front.htm}}, EBEX\footnote{EBEX: {\tt
http://groups.physics.umn.edu/cosmology/ebex/}} \citep{Oxley05},
CMBPol\footnote{CMBPol: {\tt
http://universe.nasa.gov/program/inflation.html}}).

The primordial $B$-mode anisotropy, if it exists, is at least an order
of magnitude smaller than the $E$-mode polarization. This, combined with
the difficulty of separating primordial $B$-modes from lensing $B$-modes
as well as polarized Galactic foregrounds, makes the measurement of a
potentially-existing primordial gravity wave background a great
experimental challenge. At the same time, as it is the ``smoking gun''
of inflation, the primordial $B$-mode detection and measurement can be
regarded as the ``holy grail'' of CMB measurements.

There are five main obstacles to making a measurement of the
primordial $B$-mode polarization, some being fundamental, and others
being practical complications.  The fundamental complications are: (i)
the level of the primordial signal is not guided at present by theory:
there are inflationary models consistent with the current CMB data
which predict significant levels of primordial gravity waves, as well
as models which predict negligible levels; (ii) the signal is not significantly
contaminated by lensing only on the largest scales $\ell \lap$ few
$\times 10$ where cosmic variance is important; (iii) polarized
foreground emission (mostly from our galaxy but also from
extra-galactic sources) on these scales is likely to dominate the
signal at all frequencies. Practical complications are: (iv) for the
signal to be detected in a reasonable time-scale, the instrumental
noise needs to be reduced well below the photon noise limit for
a single detector, so multiple detectors need to be used; (v)
polarized foregrounds are not yet well known, and since any detection
relies on foreground subtraction, foreground uncertainties may
seriously compromise the goal.

\citet{Tuccietal04} consider different foreground subtraction
techniques and compute the minimum detectable $r$ for future CMB
experiments for no lensing contamination.  \citet{HirataSeljak03} and
\citet{SeljakHirata04} show how the lensing $B$-mode contamination can
be subtracted out in the absence of foregrounds. Calculations of the
minimum detectable primordial gravity wave signature on the CMB
polarization data and related constraints on the inflationary paradigm
have been discussed in the literature
[e.g. \citet{KnoxSong02,SongKnox03,kinney:2003,SigurdsonCooray2005}].

\citet{Bowdenetal04} discuss the optimization of a ground-based
 experiment  (in particular, focusing on QUaD) and \citet{Huetal04}
 discuss the instrumental effects that can contaminate the detection.

In this work, we take a complementary approach. We first attempt to
forecast the performance of realistic CMB experiments: space-based,
balloon-borne and ground based; considering the covariance between
cosmological parameters and primordial parameters, effects of
foreground contamination, instrumental noise and the effect of partial
sky coverage for ground-based experiments, and then consider what
these realistic forecasts can teach us about inflation.

Given the recent advances in the knowledge of the polarized
foregrounds \citep{DASIPol, BenoitArcheops04} and in detector
technology, we then attempt here to answer the following questions:
How much can we learn about the physics of the early universe
(i.e. can we test the inflationary paradigm and constrain inflationary
models) from a realistic CMB experiment? At what point does
instrumental noise and uncertainties in the foregrounds degrade the
results that could be obtained from an ideal experiment (no noise, no
foregrounds)? What will be the limiting factor: foreground subtraction
or instrumental noise?  How much can be learned from a full-sky
space-based experiment vs. a partial-sky ground-based one? And
finally: Given the large costs involved in improving the experiment
(reducing noise by increasing the number of detectors and improving
foreground cleaning by increasing the number of frequency channels),
what is the point of diminishing returns?

We start in \S~\ref{sec:theory} by reviewing recent literature on
$B$-modes and inflationary physics, and summarizing present-day
constraints on inflationary models. We describe our forecast method in
\S~\ref{sec:method} and \S~\ref{sec:foregrounds}. In
\S~\ref{sec:experiments} we report the expected constraints on
primordial parameters from a suite of experiments including
instrumental noise and a realistic estimate of foregrounds. These
results give us some guidance in optimizing future CMB polarization
experiments for $BB$ detection and for studying inflationary physics,
which we report in \S~\ref{sec:conclusions}.  In
\S~\ref{sec:conclusions} we also draw some conclusions on the
constraints that can be applied to inflationary models based on the
capabilities of these experiments.

\section{$B$-modes and inflation} \label{sec:theory}

As \cite{peiris/etal:2003} and \cite{kinney/etal:2004} demonstrate, a
wide class of phenomenological inflation models are compatible with
the current CMB data. In the absence of specific knowledge of the
fundamental physics underlying inflation, an important consideration
is therefore to see what general statements one can make about
inflation from various types of measurements. It is well-known
(e.g. \citet{liddle/lyth:1993, copeland/etal:1993a,
copeland/etal:1993b, liddle:1994}) that a measurement of the amplitude
of the tensor modes immediately fixes the value of the Hubble
parameter $H$ during inflation when the relevant scales are leaving
the horizon, or equivalently, the potential of the scalar field
driving inflation (the {\em inflaton}) and its first derivative, $V$
and $V'$. The relations between these quantities are given by (see
e.g. \citet{lyth:1984})
\begin{eqnarray} \label{eq:inflrelations}
H &\equiv& \dot{a}/a \nonumber \\
  &\simeq& \frac{1}{M_{\rm Pl}} \sqrt{\frac{V}{3}}, \\
r &=& \frac{2 V}{3 \pi^2 M_{\rm Pl}^4 \Delta^2_{\cal R}(k_0)} \\
  &\simeq& \frac{2 V}{3 \pi^2 M_{\rm Pl}^4} \frac{1}{2.95 \times 10^{-9}
  A(k_0)} \nonumber \\
  &=& 8 M_{\rm Pl}^2 \left( \frac{V'}{V} \right)^2, 
\end{eqnarray}
where $ M_{\rm Pl} \equiv (8\pi
G)^{-1/2} = m_{\rm pl}/\sqrt{8\pi}= 2.4\times 10^{18}~{\rm GeV}$ is
the reduced Planck energy, $r=\Delta^2_h(k_0)/\Delta^2_{\cal R}(k_0)$
and $\Delta^2_h$ and $\Delta^2_{\cal R}$ are the power spectra of the
primordial tensor and curvature perturbations defined at the fiducial
wavenumber $k_0=0.002$ Mpc$^{-1}$, where the scalar amplitude $A(k_0)$
is also defined, following the notational convention of
\citep{peiris/etal:2003}.

A measurement of the spectral index of the scalar (density)
perturbations fixes $V''$, and a potential measurement of the running
of the scalar spectral index carries information about the third
derivative of the potential, $V'''$. Further, combining a measurement
of the spectral index of tensor perturbations, $n_t$, with the
tensor-to-scalar ratio $r$ allows one to test the {\em inflationary
consistency condition}, $r=-n_t/8$, that single-field slow roll models
must satisfy. However, a precise determination of the tensor spectral
index is likely to be even more difficult than a measurement of the
tensor amplitude.

The current constraint on the tensor-to-scalar ratio at 95\%
significance from WMAP data alone gives the energy scale of slow-roll
inflation as \citep{peiris/etal:2003}:
\begin{equation}
V^{1/4} \le 3.3 \times 10^{16} r^{1/4}\ \mathrm{GeV}; \label{eq:WMAP_limit}
\end{equation}
This is the familiar result that a significant contribution of tensor
modes to the CMB requires that inflation takes places at a high energy
scale. If instead of using CMB only data we use the most stringent
constraints to date on $r$, a 95\% upper limit of $r<0.36$, obtained from a compilation of
current CMB plus large scale structure data \citep{seljak/etal:2004}, one
obtains 
\begin{equation}
V^{1/4} \le 2.6 \times 10^{16}\ \mathrm{GeV}, \label{eq:Lya_limit}
\end{equation}
or equivalently, $V^{1/4} \le 2.2 \times 10^{-3}\ m_{\rm Pl}$.  
Several authors (e.g. \citet{lyth:1997, kinney:2003}) have discussed
the physical implications of a measurement of primordial $B$-mode
polarization. One can take the point of view that the inflaton $\phi$
is a fundamental field and assume the techniques of effective field
theory, for which the heavy degrees of freedom are integrated out, to
be a correct description of the physics of inflation. In this case, the
effective inflaton potential can be expanded in non-renormalizable
operators suppressed by some higher energy scale, for example the
Planck mass:
\begin{equation}
V({\phi}) = V_0 + \frac{1}{2} m^2 \phi^2 + \phi^4 \sum_{p=0}^\infty
\lambda_p \left(\frac{\phi}{m_{\rm Pl}}\right)^p. \label{eq:effptl}
\end{equation}
For the series expansion for the effective potential to be convergent
in general and the effective theory to be self-consistent, it is
required that $\phi \ll m_{\rm Pl}$. \citet{lyth:1997} showed that the
{\em width} of the potential (i.e. the distance traversed by the
inflaton during inflation) $\Delta \phi$, can also be related to
$r$. The change in the inflaton field during the $\sim 4$ e-folds when
the scales corresponding to $2 \le \ell \le 100$ are leaving the
horizon is
\begin{equation}
\frac{\Delta \phi}{m_{\rm Pl}} \sim 0.39
\left(\frac{r}{0.07}\right)^{1/2}. \label{eq:lythbound}
\end{equation}
According to this bound, high values of $r$ require changes in $\Delta
\phi$ to be of the order of $m_{\rm Pl}$; this is in fact only a lower
bound, since at least 50-60 e-folds are required before inflation ends
(e.g. \citet{liddle/leach:2003}), meaning that over the course of
inflation, $\Delta\phi$ could exceed the bound in
Eq.~\ref{eq:lythbound} by an order of magnitude.
For the present discussion, the important point is that for a
tensor-to-scalar ratio of order 0.1, the field must travel a distance
$\Delta\phi \gg m_{\rm Pl}$ over the course of inflation, so that the
effective field theory description in Eq.~\ref{eq:effptl} must break
down at some point during inflation. This fact has been used to argue,
based on self-consistency, that a very small tensor/scalar ratio is
expected in a realistic inflationary universe, leading to the
discouraging conclusion that the primordial $B$-modes in the CMB would
be unobservably small.

However, arguments based on effective field theory are not the only
way to study the dynamics of inflation. The inflaton may not be a
fundamental scalar field; in fact, as we discuss later in
\S~\ref{sec:conclusions:inf}, advances are being made building
``natural'' particle physics-motivated inflationary models based on
approximate shift symmetries and extra-dimensional setups. Also, one
can take the approach that, for the predictions of single-field
inflation to be valid, the only requirement is that the evolution of
spacetime is governed by a single order parameter. In this approach,
one can reformulate the exact dynamical equations for inflation as an
infinite hierarchy of flow equations described by the generalized
``Hubble Slow Roll'' (HSR) parameters \citep{hoffman/turner:2001,
kinney:2002, easther/kinney:2003, peiris/etal:2003,
kinney/etal:2004}. In the Hamilton-Jacobi formulation of inflationary
dynamics, one expresses the Hubble parameter directly as a function of
the field $\phi$ rather than a function of time, $H \equiv H(\phi)$,
under the assumption that $\phi$ is monotonic in time. Then the
equations of motion for the field and background are given by:
\begin{eqnarray}
\dot{\phi} &=&-2 \mpl^2 H'(\phi), \label{eq:phih}\\
\left[H'(\phi)\right]^2 -
\frac{3}{2\mpl^2}H^2(\phi)&=&-\frac{1}{2\mpl^4}V(\phi). \label{eq:hj}
\end{eqnarray}
Here, prime denotes derivatives with respect to $\phi$. Equation
(\ref{eq:hj}), referred to as the {\sl Hamilton-Jacobi Equation},
allows us to consider inflation in terms of $H(\phi)$ rather than
$V(\phi)$. The former, being a geometric quantity, describes inflation
more naturally.

This picture has the major advantage that it allows us to remove the
field from the dynamical picture altogether, and study the generic
behavior of slow roll inflation without making assumptions about the
underlying particle physics (although the underlying assumption of a
single order parameter is still present). In terms of the HSR parameters
$^{\ell}\lambda_H$, the dynamics of inflation is described by the
infinite hierarchy 
\begin{equation}
^{\ell}\lambda_H \equiv \left(\frac{m_{\rm Pl}}{4\pi}\right)^\ell
  \frac{(H')^{\ell-1}}{H^\ell} \frac{d^{(\ell+1)} H}{d\phi^{(\ell+1)}}.
  \label{eq:hier}
\end{equation}
The flow equations allow us to consider the model space spanned by 
inflation using Monte-Carlo techniques. Since the dynamics are
governed by a set of first-order differential equations, one can
specify the entire cosmological evolution by choosing values for the
slow-roll parameters $^{\ell}\lambda_H$, which completely specifies the
inflationary model. 

Note, however, two caveats: first, in practice one
has to truncate the infinite hierarchy at some finite order; in this
paper we retain terms up to 10th order. Secondly, the choice of slow
roll parameters for the Monte-Carlo process necessitates the assumption
of some priors for the ranges of values taken by the
$^{\ell}\lambda_H$. In the absence of any {\sl a priori} theoretical
knowledge about these priors, one can assume flat priors with some
ranges dictated by current observational limits, and the requirement
that the potential satisfies the slow-roll conditions. Changing this
``initial metric'' of slow-roll parameters {\sl changes} the
clustering of phase points on the resulting observational plane of a
given Monte-Carlo simulation. Thus, {\sl the results from these
  simulations cannot be interpreted in a statistical way}.

Nevertheless, important conclusions can be drawn from the results of
  such simulations (e.g. \citet{kinney:2002, peiris/etal:2003}); they
  show that models do not cover the observable parameter space
  uniformly, but instead cluster around certain attractor regions. And
  these regions do not correspond especially closely to the
  expectations from effective field theory; they show significant
  concentrations of points with significant tensor/scalar ratio.

Recent work by \citet{efstathiou/mack:2005} shows that, in the
inflationary flow picture, there is no well-defined relation of the
form Eq.~\ref{eq:lythbound}. However, a reformulation of the bound
in Eq.~\ref{eq:lythbound} is found if one imposes observational
constraints on the scalar spectral index $n_s$ and its running with
scale $d n_s/d\ln k$. 

The precise values and error bars on these parameters depend on the
particular data set (or combinations of data sets) used in the
parameter estimation, and some data sets yield cleaner and more
reliable determinations than others
\citep{SeljakMcdonaldMakarov03}. Nevertheless, for our purposes  we
shall use the tightest constraints available at present. These come
from a compilation of CMB data, and large scale structure data from
galaxy surveys and Lyman-$\alpha$ forest lines
(e.g. \citet{seljak/etal:2004}): at 95\% confidence these limits are
$0.92 < n_s < 1.06$ and $-0.04 < dn_s/d\ln k < 0.03$.

Here we illustrate how an empirical relation between $\Delta\phi$ and
$r$ can be found and show that this relation is somewhat insensitive
to the priors on the $^{\ell} \lambda_H$ parameters.

\begin{figure}
\includegraphics[scale=0.35]{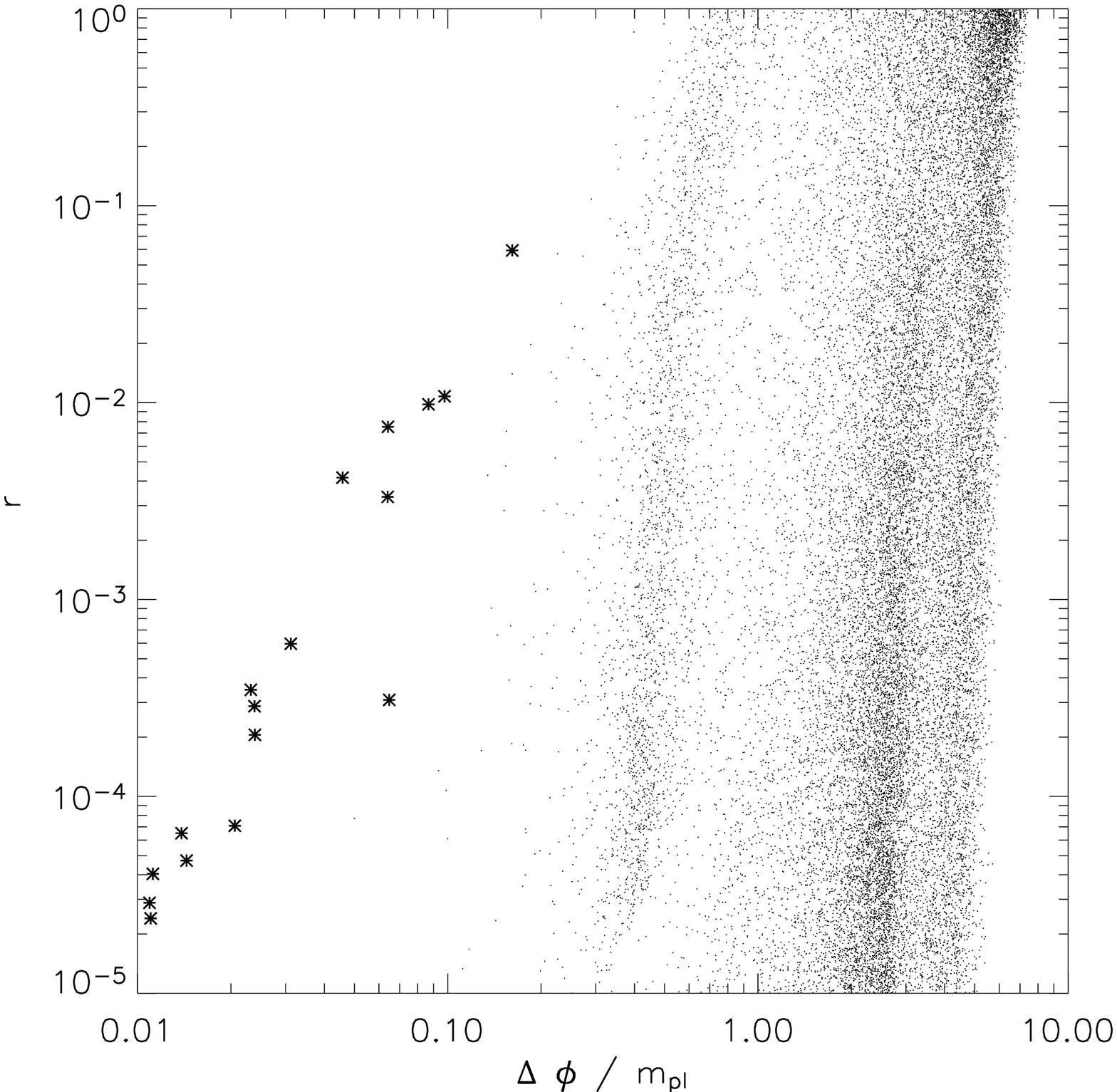}
\includegraphics[scale=0.35]{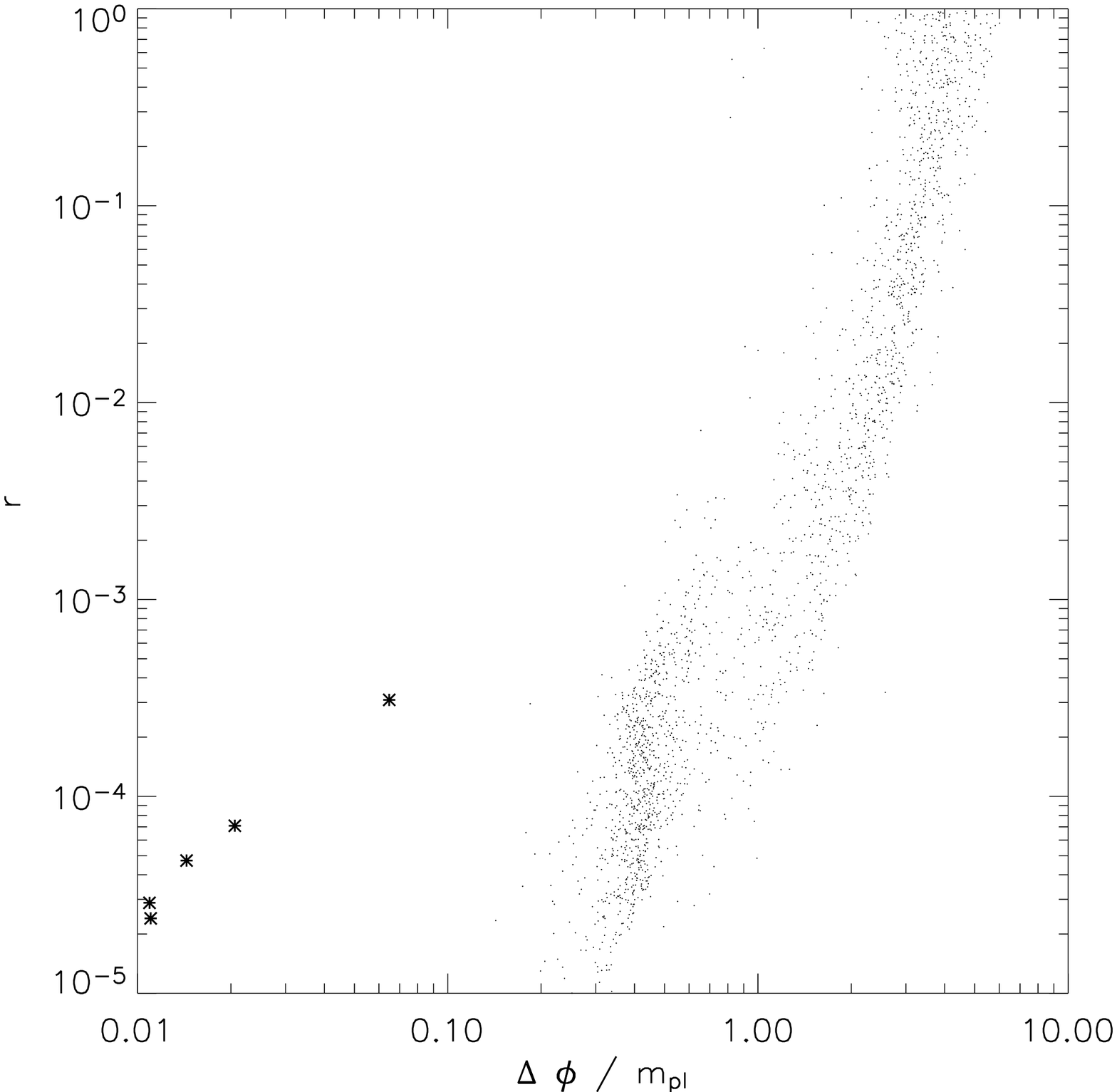}
\caption{The absolute value of the range traversed by the scalar field
  $\phi$ over the final 55 e-folds of inflation, $\Delta \phi$,
  plotted against the tensor-to-scalar ratio $r$: (top) all models
  that sustain at least 55 e-folds of inflation in a 2-million point
  Monte-Carlo simulation to 10th order of the inflationary flow
  equations; (bottom) the subset of models that satisfy current
  observational constraints on $n_s$ and $d n_s / d \ln k$, as
  discussed in the text. Hybrid models, which have $n_s>1$, are shown
  as stars, and the  rest as dots (see text). This simulation has a
  wider prior on $^4\lambda_H$ and higher order slow roll parameters.}
\label{fig.lythboundwmap}
\end{figure}

\begin{figure}
\includegraphics[scale=0.35]{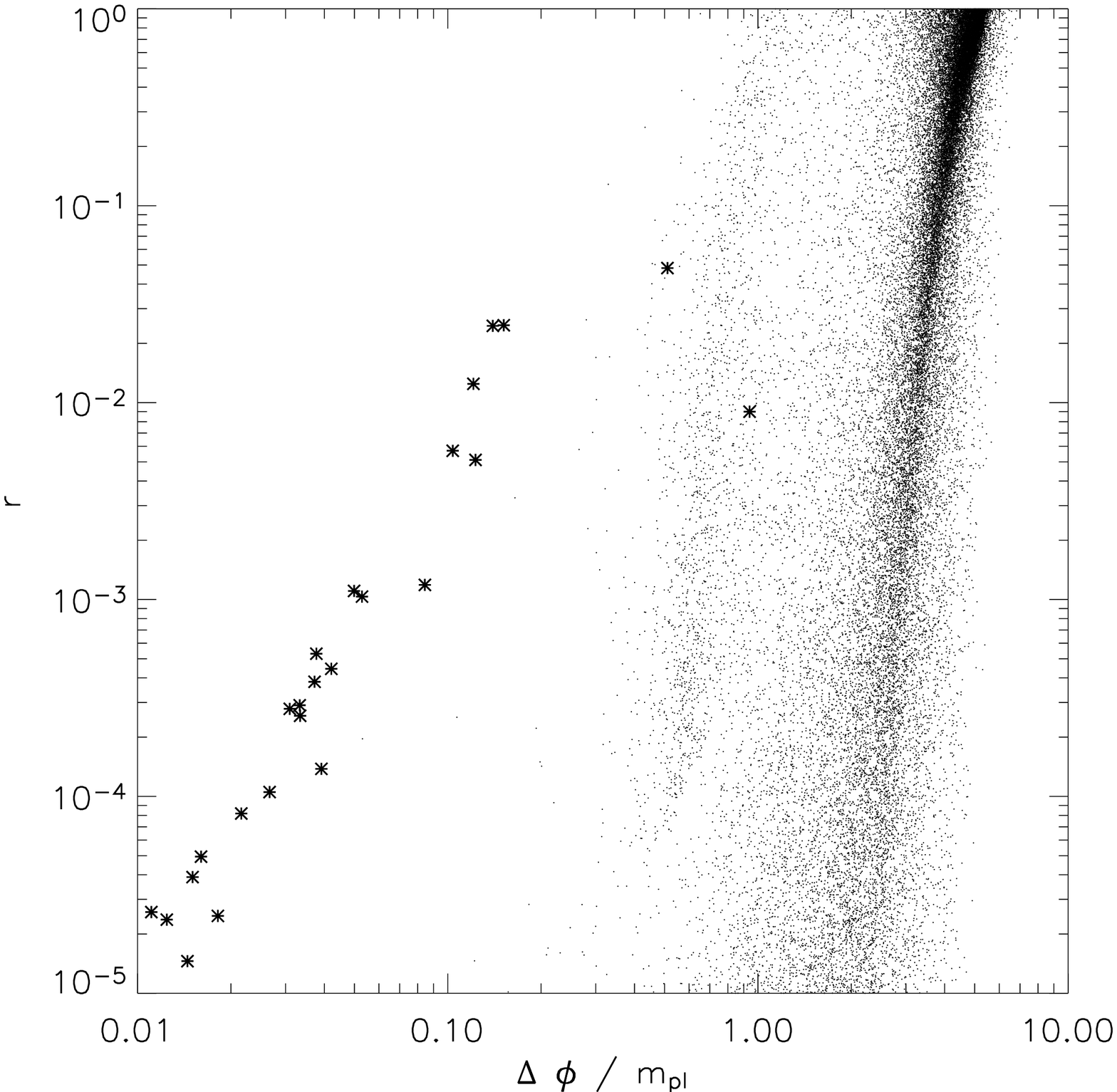}
\includegraphics[scale=0.35]{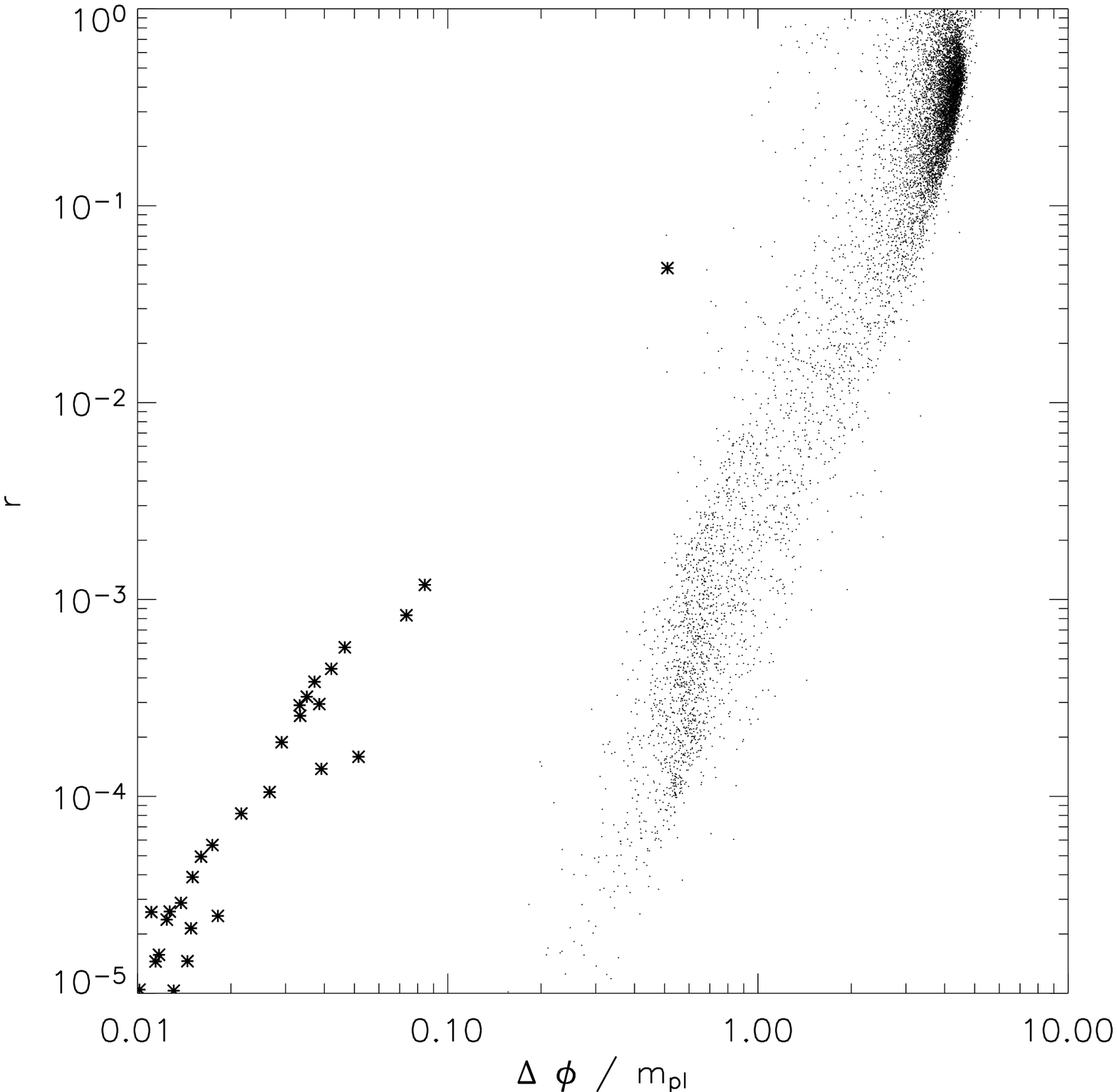}
\caption{Otherwise identical to Fig.~\ref{fig.lythboundwmap}, except
  that the simulation shown here has a narrower prior on $^4\lambda_H$
  and higher order slow roll parameters.} 
\label{fig.lythboundalldata}
\end{figure}

Figures \ref{fig.lythboundwmap} and \ref{fig.lythboundalldata} show
the absolute value of $\Delta \phi$ over the final 55 e-folds of
inflation, plotted against the tensor-to-scalar ratio
$r$. Fig. \ref{fig.lythboundwmap} shows a 2-million point Monte-Carlo
simulation of the flow equations, allowing the fourth slow roll
parameter, $^4\lambda_H$, to have a wide prior $^4\lambda_H =
[-0.025,0.025]$. This prior on $^4\lambda_H$ has been (arbitrarily)
set to have the same width as the bound on $^3\lambda_H$ allowed by
the WMAP data \citep{peiris/etal:2003}. A wide prior is also allowed
on the higher order HSR parameters. In the simulation shown in
Fig. \ref{fig.lythboundalldata}, the width of the prior on
$^4\lambda_H$ has been decreased by a factor of five, $^4\lambda_H =
[-0.005,0.005]$. The higher order HSR parameters also have a narrower
prior than the previous case.  These choices are motivated by  the
prejudice that higher-order parameters in the expansion should have
narrower priors around zero, and by the fact that tighter constraints
on $^3\lambda_H$ are possible from CMB and large-scale structure data
compilations or future higher-precision CMB data.

As was emphasized earlier, the top panels of these figures show that
the clustering of points in the two cases is different due to the
different initial distributions allowed. The bottom panels show the
remaining points after the observational constraints from the above
compilation of CMB and large scale structure have been applied. It is
seen that with this prior in phase-space, there is now a fairly 
well-defined relation  between $\Delta\phi$ and $r$, and that even
though the distributions
in the two simulations with initial priors of different width are
different, this relationship is roughly the same in both cases. As
reported in \citet{efstathiou/mack:2005}, this reformulation of the
bound in Eq.~\ref{eq:lythbound} is given by
\begin{equation}
\frac{\Delta \phi}{m_{\rm Pl}} \simeq 6 r^{1/4}. \label{eq:lythflow}
\end{equation}

As \citet{efstathiou/mack:2005} point out, the relation in
Eq.~\ref{eq:lythflow} is so steep that, to probe models with small
field variations where an effective field theory description is likely
to be valid, one needs to be able to detect $r \le
10^{-4}$. Conversely, a detection of a larger tensor component would
support a large-field model for inflation. 

Note that the discussion and the bound stated in Eq. \ref{eq:lythflow}
apply to non-hybrid models, plotted as dots. Figures
\ref{fig.lythboundwmap} and \ref{fig.lythboundalldata} also show
Monte-Carlo simulations which qualify as hybrid inflation models
(e.g. \citet{Linde:1993cn}), plotted as stars. These indeed show a few
small-field models which have significant values of $r$.  However, it
must be noted that these models inflate forever till inflation is
suddenly ended at some critical time by a second field; the
observables for the Monte-Carlo simulations for models in this class
have been computed arbitrarily at 400 e-folds before the end of
inflation for the purposes of these figures - observables computed at
a different time during inflation would obviously yield different
results. If a value of $r \gap 10^{-4}$ is detected, hybrid models can
be easily distinguished from large-field models by looking at the
scalar spectral index: large-field models predict a red spectral index
while hybrid models predict a blue index.

We will now try to assess the minimum level of $r$ detectable with
an idealized and a realistic CMB experiment.

\section{Method}\label{sec:method}
Here we describe the methods and the notation we will use for our
error forecasts. 
\subsection{Definitions and Conventions}

Following \citet{BJK}, we define
\begin{equation}
C^{XY}_{\ell}=\frac{\ell (\ell+1)}{2\pi}\sum_{m}\frac{a^{X*}_{\ell m}
a^Y_{\ell m}}{2\ell+1},
\end{equation}
where $a_{\ell m}$ denotes the spherical harmonic coefficients, the $XY$
superscript can be $TT, TE, EE$ or $BB$, $a^X_{\ell m}$ denotes the
spherical harmonic coefficients  of the $X$ signal and
\begin{equation}
\Cc_{\ell}=\frac{\ell (\ell+1)}{2\pi}C_{\ell}\,.
\end{equation}

For our forecast calculations, we will assume Gaussian beams: if
$\Theta_{FWHM}$ denotes the FWHM of a Gaussian beam, then $\sigma_b=0.425
\ \Theta_{FWHM}$.

The noise per multipole $n_0$ is given by
$n_0=\sigma^2_{pix}\Omega_{pix}$ where
$\Omega_{pix}=\Theta_{FWHM}\times \Theta_{FWHM}=4\pi f_{sky}/N_{pix}$
is the pixel (beam) solid angle and $N_{pix}$ is the number of pixels
(independent beams) in the survey region and $f_{sky}$ is the fraction
of the sky covered by observations. $\sigma^2_{pix}$ is the variance
per pixel, which can be obtained from the detector sensitivity $s$ as
$\sigma_{pix}=s/\sqrt{Nt}$. $N$ is the number of detectors  and
$t$ the integration time per pixel.  With these conventions  the
noise bias becomes ${\cal N}_{\ell}=\frac{\ell
(\ell+1)}{2\pi}n_0\ e^{\ell^2\sigma_b^2}$.

As the receiver temperature $T_{rec}$ is limited to be
$T_{rec}>h\nu/k$ ($k$ is the Boltzmann constant and $h$ is the Planck
constant), the r.m.s temperature $\Delta T_{rms}$ is limited to be
$\Delta T_{rms}>(T_{CMB}+T_{rec})/\sqrt{\Delta \nu}{\Delta t}$ where
$\Delta \nu$ is the bandwidth and $\Delta t$ the integration time.
This limit can be overcome  by using multiple detectors; for example
QUIET and PolarBeaR will have of the order of $10^3$ detectors. 
\subsection{Likelihood}

Assuming that CMB multipoles are Gaussian-distributed, the likelihood
function for an ideal, noiseless, full sky  experiment  can be written
as
\begin{eqnarray}
-2\ln{\cal L}&=&\sum_{\ell} (2\ell+1)\left\{
 \ln\left(\frac{C_{\ell}^{BB}}{\hat{C}_{\ell}^{BB}}
 \right)\right.\nonumber \\
 & + & \ln\left(\frac{C_{\ell}^{TT}C_{\ell}^{EE}-(C_{\ell}^{TE})^2}{\hat{C}_{\ell}^{TT}\hat{C}_{\ell}^{EE}-(\hat{C}_{\ell}^{TE})^2}\right) \nonumber \\
&+&
 \frac{\hat{C}_{\ell}^{TT}C_{\ell}^{EE}+C_{\ell}^{TT}\hat{C}_{\ell}^{EE}-2\hat{C}_{\ell}^{TE}C_{\ell}^{TE}}{C_{\ell}^{TT}C_{\ell}^{EE}-(C_{\ell}^{TE})^2}\nonumber \\
 &+& \left. \frac{\hat{C}_{\ell}^{BB}}{C_{\ell}^{BB}}-3\right\}, \label{eq:like_ideal}
\end{eqnarray}
where $\hat{C}^{XY}_{\ell}$ is the estimator for the measured angular
 power spectrum. The likelihood (\ref{eq:like_ideal}) has been
 normalized  with respect to the maximum likelihood, where
 $C^{XY}_{\ell}=\hat{C}^{XY}_{\ell}$.

In the case of a more realistic experiment (with partial sky coverage,
and noisy data) we obtain:
\begin{eqnarray}
-2\ln{\cal L}&=&\sum_{\ell} (2\ell+1)\left\{
 f_{sky}^{BB}\ln\left(\frac{\CC_{\ell}^{BB}}{\hat{\CC}_{\ell}^{BB}}
 \right)\right.\nonumber \\
 & + & \sqrt{f_{sky}^{TT}f_{sky}^{EE}}\ln\left(\frac{\CC_{\ell}^{TT}\CC_{\ell}^{EE}-(\CC_{\ell}^{TE})^2}{\hat{\CC}_{\ell}^{TT}\hat{\CC}_{\ell}^{EE}-(\hat{\CC}_{\ell}^{TE})^2}\right) \nonumber
 \\
&+&\sqrt{f_{sky}^{TT}f_{sky}^{EE}}\frac{\hat{\CC}_{\ell}^{TT}\CC_{\ell}^{EE}+\CC_{\ell}^{TT}\hat{\CC}_{\ell}^{EE}-2\hat{\CC}_{\ell}^{TE}\CC_{\ell}^{TE}}{\CC_{\ell}^{TT}\CC_{\ell}^{EE}-(\CC_{\ell}^{TE})^2}\nonumber \\
& +& \left. f_{sky}^{BB}\frac{\hat{\CC}_{\ell}^{BB}}{\CC_{\ell}^{BB}}-2\sqrt{f_{sky}^{TT}f_{sky}^{EE}}-f_{sky}^{BB}\right\}
\label{eq:like_real}
\end{eqnarray}
In Eq. \ref{eq:like_real}, $\CC^{XY}_{\ell}=\Cc^{XY}_{\ell}+{\cal
N}^{XY}_{\ell}$ where ${\cal N}_{\ell}$ denotes the noise bias,
$\hat{\CC}^{XY}$ denotes the measured angular power spectrum which
includes a noise bias contribution, $f_{sky}^{XY}$ denotes the
fraction of sky observed, $X,Y=\{T,E,B\}$ and we have allowed
different fraction of the sky to be observed in $T$, $E$ and $B$.

Note that this equation accounts for the effective number of modes
allowed in a partial sky, but does not account for the mode
correlation introduced by the sky cut which may smooth power spectrum
features. The details of the mode-correlation depend on the specific
details of the mask, but for $\ell$ greater than the characteristic
size of the survey our approximation should hold.

The expression for the likelihood (\ref{eq:like_real}) is only valid
for a single frequency experiment.  The exact expression for
multi-frequency experiments which takes into account for both auto-
and cross- channels power spectra cannot be written in a
straightforward manner in this form. However, for $N_{chan}$
frequencies with the same noise level, considering both auto and cross
power spectra is equivalent to having one effective frequency with an
effective noise power spectrum lower by a factor $N_{chan}$.  This can
be understood in two ways. If one were to combine $N_{chan}$
independent maps, the resulting map will have a noise level lower by a
factor $\sqrt{N_{chan}}$; thus the noise power spectrum will be that
of an auto-power spectrum, reduced by a factor ${N_{chan}}$.  One
could instead compute auto and cross $C_{\ell}$ and then combine them
in an minimum variance fashion.  Let us assume Gaussianity for
simplicity. When combining the $C_{\ell}$, the covariance matrix in the
absence of cosmic variance will be:
\begin{equation}
\Sigma_{i i' j j'}=\frac{1}{2l+1}(n^i n^j \delta_{ij} \times  n^{i'}
n^{j'}\delta_{i'j'}+n^i n^{j'} \delta_{ij'}\times n^{i'} n^j
\delta_{i'j})
\end{equation}
where $i,j$ run through the channels, $n^in^j=N^{ij}$ and
$\delta_{ij}$ denotes the Kronecker delta. Note that this has been
computed for the no cosmic variance case, as there is no cosmic
variance involved when combining $C_{\ell}$'s computed from  the same
sky at different wavelengths).  Thus, if $N_{eff}$ is the same for all
channels then the covariance between, say, $C_{\ell,\ VV}$ \& $C_{\ell,\
VV}$ is twice that between say  $C_{\ell,\ VW}$ \& $C_{\ell,\ VW}$; the
covariance between $C_{\ell,\ VV}$ \& $C_{\ell,\ WW}$ and $C_{\ell,\
VW}$ \& $C_{\ell,\ QV}$ etc. is zero. The minimum variance combination
will thus yield an effective noise power spectrum of $N/N_{chan}$.
 
 We can generalize the above considerations to the case where  the
 different channels have different noise levels. This is useful for
 example in the case where each channel has a different residual
 foreground level,  which, as we will show below, can be approximated
 as a white noise contribution in addition to the instrumental noise.
 The optimal channel combination would be:
\begin{equation}
C_{l}=\frac{\sum_{i,j,j\ge i} w_{i,j} C^{i,j}_l}{\sum_{i,j} w_{i,j}}
\end{equation}
where $w_{i,j}$ are weights and are given by $[N_i N_j
1/2(1+\delta_{ij})]^{-1}$. The resulting noise will be given by
\begin{equation}
N_{eff}=\left(\sum_{i,j,j\ge i}
\frac{1}{N_{eff,i}N_{eff,j}\frac{1}{2}(1+\delta_{ij})}\right)^{-1/2}
\end{equation}

\subsection{Error calculation}

The Fisher information matrix is defined as \citep{Fisher35}:
\begin{equation}
F_{ij}=\left \langle -\frac{\partial^2 L}{\partial \alpha_i\alpha_j}
\right \rangle|_{\alpha=\bar{\alpha}}
\end{equation}
where $L\equiv \ln {\cal L}$ and $\alpha_i$, $\alpha_j$ denote model
parameters. The Cramer-Rao inequality says that the minimum standard
deviation, if all parameters are estimated from the same data,   is
$\sigma_{\alpha_i}\ge (F^{-1})_{ii}^{1/2}$ and that in the limit of a
large data set, this inequality becomes an equality for the maximum
likelihood estimator.

For our error forecast we use the Fisher matrix approach. Thus here,
${\bf \alpha}$ denotes a vector of cosmological parameters (we
consider $\{r,n,n_t,dn/d\ln k, Z ,\Omega_b h^2 \equiv w_b, \Omega_c
h^2 \equiv w_c, h, \Omega_k, A(k_0)\}$, where $Z\equiv\exp(-2 \tau)$
), $\bar{{\bf \alpha}}$ denotes the fiducial set of cosmological
parameters, $\alpha_i$ being the $i^{th}$ component of that vector,
and ${\cal L}$ is given by Eq. \ref{eq:like_real}.  For the fiducial
parameters, we will consider three cases for $r$, $r=\{0.01,0.03,
0.1\}$ and  $r=0.001$, $r=0.0001$  as a special cases; for the other
cosmological parameters we use $n=1$, $n_t=-r/8$, $d\ln n/d\ln k=0$,
$\exp(-2 \tau)=0.72$ (corresponding to $\tau=0.164$), $\Omega_b
h^2=0.024$, $\Omega_c h^2=0.12$, $h=0.72$, $\Omega_k= 0$,
$A(k_0)=0.9$, in agreement with WMAP first year results
\citep{SpergelWMAP03}.

Following \citet{SpergelWMAP03}, the pivot point for $n$ and $A$ is at
$k=0.05$ Mpc$^{-1}$ and the tensor to scalar ratio $r$ is also defined
at $k=0.05$ Mpc$^{-1}$. Making use of the consistency condition and
assuming no running of the scalar or tensor spectral indices, $r(k_0)$
can be related to the $r$ defined at any other scale $k_1$ by
\begin{equation}
r(k_1) = r(k_0) \left(\frac{k_1}{k_0}\right)^{-r(k_0)/8 + 1 -
  n_s(k_0)} \label{eq:rscaling}.
\end{equation}
Therefore, for our choice of a scale-invariant scalar power spectrum,
e.g. $r(0.05) = 0.1$ corresponds to $r(0.002) = 0.104$ and our results are
thus not significantly dependent on the choice of the pivot point.

Note that for a fixed value of $r$, the height of the $BB$
``reionization bump'' depends on $\tau$ approximately as $\tau^2$, but
the position of the ``bump'' maximum also varies. The $\tau$
determination from WMAP has a 1-$\sigma$ error of $\sim 0.07$
\citep{SpergelWMAP03}. Thus for a fiducial value of $\tau \sim 0.1$,
one sigma below the best fit, the detectability of $r=0.03$ ($r=0.1$)
from $\ell \lap 20$ is roughly\footnote{Note that since the ``bump''
location changes as $\tau$ changes this argument is only qualitative.}
equivalent to the detectability of $r=0.01$ ($r=0.03$) in the best fit
model. On the other hand, for a fiducial value of $\tau$ one sigma
above the best fit, the detectability of $r=0.01$ ($r=0.03$) from
$\ell \lap 20$ is equivalent to the detectability of $r=0.03$
($r=0.1$) in the best fit model. This consideration affects only large
sky coverage experiments which probe the reionization bump in the $BB$
spectrum at $\ell <10$, and does not affect the smaller scale ground
based experiments. We therefore only calculate the case of the
one-sigma lower bound fiducial $\tau=0.1$ for the space-based test
case with $r=0.001$ and the balloon-borne large angle experiment SPIDER.
As a higher value for $\tau$ boosts the $BB$ signal, we investigate
this lower value of $\tau$ in order to be conservative in our predictions.

We use all the power spectrum combinations as specified in
Eqs. \ref{eq:like_ideal} and \ref{eq:like_real}, computed to
$\ell=1500$ and using the experimental characteristics specific to
each experiment we study.  For the partial sky experiments we add WMAP
priors on $n$, $dn/d\ln k$, $Z$, $\Omega_b h^2$, $\Omega_c h^2$, $h$,
and $A$.

We use the ``three point rule'' to compute the second derivative of
$L$ when $i=j$:
\begin{equation}
\frac{\partial^2L}{\partial^2\alpha_i}=\frac{L(\bar{\alpha}_{-\delta\alpha_i})-2L(\bar{\alpha})+L(\bar{\alpha}_{+\delta\alpha_i})}{\delta\alpha_i^2}.
\end{equation}
Here, by $L(\bar{\alpha})$ we mean $L$ evaluated for the fiducial set
of cosmological parameters. $\bar{\alpha}_{+\delta\alpha_i}$
denotes the vector of cosmological parameters made by the fiducial values for
all elements except the $i^{th}$ element, which is
$\bar{\alpha}_i+\delta \alpha_i$;  $\delta\alpha_i$ is a small change
in $\alpha_i$. 

For $i \ne j$ we use:
\begin{eqnarray}
\frac{\partial^2L}{\partial\alpha_i \partial
\alpha_j}&\!\!\!\!\!\!\! = &\!\!\!\!\!\!
\left\{[L(\bar{\alpha}_{+\delta\alpha_i+\delta\alpha_j})-L(\bar{\alpha}_{+\delta\alpha_i-\delta\alpha_j})]\right.\nonumber
\\
&\!\!\!\!\!\!\-&\!\!\!\!\!\!\!\left.[L(\bar{\alpha}_{-\delta\alpha_i+\delta\alpha_j})-L(\bar{\alpha}_{-\delta\alpha_i-\delta\alpha_j})]\right\}\!/(2\delta\alpha_i
2 \delta\alpha_j)
\end{eqnarray}

where $\bar{\alpha}_{+\delta\alpha_i-\delta\alpha_j}$ denotes a
vector of cosmological parameters where all components are equal to
the fiducial cosmological parameters, except components $i$ and $j$
which are $\bar{\alpha}_i+\delta\alpha_i$ and
$\bar{\alpha}_j-\delta\alpha_j$ respectively.

For high multipoles the Central Limit Theorem ensures that the
likelihood is well approximated by a Gaussian. Thus, if one also
neglects the coupling  between $TT$, $TE$ and $EE$, the high $\ell$ Fisher
matrix can be well approximated by
\begin{equation}
-2 \langle \frac{\partial \ln {\cal L}}{\partial \alpha_i \partial
 \alpha_j}\rangle|_{\alpha=\bar{\alpha}}=\sum_{P,\ell}
 \frac{\partial{C^{P}_{\ell}}}{\partial \alpha_i}\Sigma^{-1}_{\ell
 \ell'}\frac{\partial C^{P}_{\ell'}}{\partial \alpha_j},
\end{equation}
where $P={TT, EE, BB, TE}$, and $\Sigma$ denotes the covariance
matrix. Therefore the Fisher matrix calculation can be sped up greatly
by precomputing  $\frac{\partial \Cc^P_{\ell}} {\partial \alpha_i}$.
However, since most of the $r$ signal comes from low $\ell$, we will not
use this approximation here.

\subsection{Detection vs Measurement}

There are two different approaches in reporting an experiment's
capability to constrain $r$. In the first approach, one considers the
null hypothesis of zero signal (i.e. $r=0$); then one asks with what
significance a non-zero value of $r$ could be distinguished from the
null hypothesis. This approach gives the statistical significance of a
detection, but it does not yield a measurement of $r$. To obtain a
measurement of $r$ with an error-bar, a different approach is needed.
In this case, the value of $r$ is obtained via a Bayesian maximum
likelihood analysis, and the cosmic variance contribution for non-zero
$r$  is included in the error calculation. Since a measurement of $r$
is needed to constrain inflationary models, here we follow the second
approach. Thus we do not report the minimum value of $r$ that can be
distinguished from zero, but error-bars for several fiducial $r$ values.


\section{Foregrounds} \label{sec:foregrounds}

As anticipated in the introduction, foregrounds are one of the main
obstacles to the detection and measurement of primordial $B$-modes: they
are likely to dominate the signal at all frequencies. And, while
foreground intensities are relatively well known, polarized
foregrounds are not. Here we will neglect the effect of foregrounds on
the CMB temperature, as foreground contamination in the temperature
data at the angular scales considered is much smaller than the
contamination to the polarized signal and foreground cleaning is
expected to  leaves a negligible contribution in the temperature
signal \citep{BennettWMAP03a}.

CMB polarized foregrounds arise due to free-free, synchrotron, and
dust emission, as well as due to extra-galactic sources such as radio
sources and dusty galaxies. Here, we will consider only synchrotron
and dust emission; in fact, free-free emission is expected to be
negligibly polarized, and the radio source contribution remaining
after masking the sky for all sources brighter than $\sim $1 Jy is
always well below other foregrounds at large angular scales.  E.g.
\citet{Tuccietal04}) find that this residual point source
contribution is $\lap 1$\% of the other foregrounds in the power for
$\ell \lap 200$. Note, however, that at $\ell \gap 100$, this
contribution that we neglect can be as important in the BB power
spectrum as the lensing signal (see \citet{Tuccietal04b}), and
therefore can seriously hinder the delensing implementation.

 We also neglect the contribution from dusty galaxies.

\subsection{Synchrotron}
Free electrons spiral around the Galactic magnetic field lines and
emit synchrotron radiation. This emission can be up to 75\% polarized,
and is the main CMB foreground at low frequencies.  We model the power
spectrum of synchrotron as (e.g. \citet{tegmarketal00, deoliveiracosta03}):
\begin{equation}
C^{S,XY}_{\ell}=A^{S}\left(\frac{\nu}{\nu_0}\right)^{2
\alpha_S}\left(\frac{\ell}{\ell_0}\right)^{\beta_S}
\label{eq:synCl}
\end{equation}
where we have assumed that Galactic synchrotron emission has the same
amplitude for $E$ and $B$.  The unpolarized synchrotron intensity has
$\alpha_S\sim -3$ with variations across the sky of about $\Delta
\alpha_S\sim 0.15$ (e.g. \citet{Plataniaetal98, Plataniaetal03,
BennettWMAP03a}) and $\beta_S \simeq -3$ (e.g. \citet{Giardino02} and
references therein) with variation across the sky
\footnote{At low Galactic latitudes, the synchrotron index shows a
flattening, but we assume here that the mask will exclude those
regions}. Information on polarized synchrotron is limited at present
to frequencies much lower than CMB frequencies (e.g.
\citet{Carrettietal2005a, Carrettietal2005b}  and mostly low Galactic
latitudes).

Recent estimates for $\beta_S$ for polarized synchrotron emission
\citep{Tuccietal02, Bruscolietal02} show spatial and frequency
variations.  For example, \citet{Bruscolietal02} study the angular
power spectrum of polarized Galactic synchrotron at several
frequencies below 3 GHz and at several Galactic latitudes, and find a
wide variation mostly due to Faraday rotation and variations in the
local properties of the foregrounds. \citet{Bernardietal04a} find that
at high galactic latitudes there is  no significant dependence of the
slope on frequency. Here we consider $\beta_S=-1.8$ for both $E$ and
$B$, consistent with e.g.  \citet{Baccigalupietal01,Bruscolietal02,
  Bernardietal04a, Bernardietal03}.

We take the reference frequency of $\nu_0=30$ GHz and the reference
multipole $\ell_0=300$ so that we can use the DASI 95\% upper limit
\citep{DASIPol} of 0.91 $\mu$K$^2$ to set the amplitude $A_S$.  Since the
normalization is set at $\ell =300$ and most of the $r$ signal is at
$\ell <100$, a choice of a flatter  $\beta_S$ would have the effect of
reducing the synchrotron contamination at low $\ell$.

 We will consider three cases: an amplitude equal to the DASI limit
(pessimistic case)  and half of that amplitude (reasonable case).
Recent results by \citet{Carrettietal2005b} show a detection of
synchrotron polarized emission in a clean area of the sky at 2.3
GHz. The amplitude of this signal is about an order of magnitude below
the DASI upper limit. We will consider this amplitude (10\% of the
DASI limit) as a minimum amplitude (optimistic case) achievable only
for partial sky experiments which can look at a clean patch of the
sky. Unless otherwise stated we will assume the pessimistic case.

\subsection{Dust} 
We ignore a possible anomalous dust contribution
 (e.g. \citet{Finkbeiner:2004je, DraineLazarian98}) as it is important
 only at frequencies below 35 GHz or so \citep{LazarianDraine03}.  We
 model the dust signal as:
\begin{equation}
C^{D,XY}_{\ell}=
p^2A^D\left(\frac{\nu}{\nu_0}\right)^{2\alpha_D}\left(\frac{\ell}{\ell_0}\right)^{\beta^{XY}_D}\left[\frac{e^{h\nu_0/KT}-1}{e^{h\nu/KT}-1}\right]^2
\label{eq:dustCl}
\end{equation}
assuming the temperature of the dust grains to be constant across the
sky, $T=18$K. We take $\alpha_D=2.2$ \citep{BennettWMAP03a}, and set
our reference frequency to be $\nu_0=94$ GHz. The amplitude $A^D$ is
given by the intensity normalization of \citet{FDS99}, extrapolated to
$94$ GHz, and $p$ is the polarization fraction. Since we want to use
the  Archeops upper limit for the diffuse dust component \citep{BenoitArcheops04} of $p =$ 5\% at
$\ell=900$ (roughly corresponding to the resolution in
\citet{BenoitArcheops04}), we set $\ell_0=900$. Archeops finds
$\alpha_D=1.7$; with our choice the extrapolation of the dust
contribution at higher frequency is slightly more conservative.
Since a weak Galactic magnetic field of 3 $\mu$G already gives a 1\%
\citep{Padoanetal01} polarization, the polarization fraction should be
bound to be  between 1--5\%.  Unless otherwise stated, we will assume
a 5\% polarization fraction.  For $C_{\ell}^{XY,D}$ we follow
\citet{PrunetLazarian02, Prunetetal98}: $\beta^{EE}_D=-1.3$,
$\beta^{BB}_D=-1.4$, $\beta^{TE}_D=-1.95$, $\beta^{TT}_D=-2.6$ (in
agreement with the measurement of starlight polarization of
\citet{FosalbaLazarianPrunet02}).

As parameters like $\beta^{XY}_D$, $\beta^{XY}_S$, $\alpha_D$,
$\alpha_S$ and $p$ may show spatial variations, the optimal frequency
band may be different for full sky and partial sky experiments and may
vary between different patches of the sky.
  
\subsection{Propagation of foreground subtraction errors}

The foreground treatment presented above is quite simplistic and
cannot capture the foreground properties completely. However, we will
not be using these estimates to subtract the foreground from the
signal.  Instead, we assume that foreground subtraction can be done
correctly down to a given level (e.g. 1\%, 10\%), and use these
foreground models to propagate the effects of foreground subtraction
residuals into the resulting error-bars for the cosmological
parameters. In other words the modeling enters in the error-bars, not
in the signal. Therefore, as long as the foreground assumptions are
reasonable, and foregrounds can be subtracted at the assumed level,
the results are relatively insensitive to the details of the
foreground model. Of course, one cannot know in advance whether the
foreground will be subtracted at that level. For this reason we
considered several different options (10\%, 1\% etc..).

To propagate foreground subtraction errors we proceed as follows. We
assume that foregrounds are subtracted from the maps via foreground
templates or using multi-frequency information with techniques such as
MEMs (e.g. \citet{BennettWMAP03a}), ICA
\citep{Baccigalupietal04,Stivolietal05}. We write the effect on the
power spectrum of the residual Galactic contamination as an additional
``noise-like'' component $C_{\ell}^{res,fg,XY}$ composed by a term
proportional to the foreground $C_{\ell}$ and a noise term:
\begin{eqnarray}
C_{\ell}^{res,fg,XY}(\nu)&=&C_{\ell}^{fg,XY}(\nu)\sigma^{fg,XY}\nonumber
\\
&+&N^{fg,XY}\left(\frac{\nu}{\nu_{tmpl}}\right)^{-2\langle
\alpha_{fg}\rangle}
\label{eq:clres}
\end{eqnarray}
where $fg$ denotes dust (D) or synchrotron (S),  $\langle
\alpha_{fg}\rangle$ denotes the average value of the spectral index,
and $\sigma^{fg,XY}$ quantifies how good the foreground subtraction is
(e.g. $\sigma^{fg,XY}$=0.01 for a 1\% residual in the $C_{\ell}$).
$C^{fg,XY}_{\ell}$ is given by Eq. \ref{eq:synCl} for synchrotron and
Eq. \ref{eq:dustCl} for dust; $N^{fg,XY}$ denotes the noise power spectrum
of the template map.  As an estimate, we assume that the noise
spectrum in the template maps is a white noise equal to the noise
spectrum of one of the channels reduced by a factor $ 1/2 \times n
(n-1)/2$. The factor of 1/2 arises from the fact that foreground
templates are effectively obtained by subtracting maps at two
different frequencies (which increases the noise in the map by a
factor $\sim \sqrt{2}$), and that there are $n(n-1)/2$ map
pairs. Although the map pairs are not all independent, we expect that
experiments will have some channels both at higher and lower
frequencies than those devoted to cosmology, to be used for
foreground studies. We thus consider this a reasonable estimate for
the template noise.
  
$\nu_{tmpl}$ denotes the frequency at which the template is
created. For this forecast we assume that $\nu_{tmpl}$  coincides with
the frequency of the experiment channel where the foreground
contamination is  highest.  To justify the {\it ansatz} Eq. (\ref{eq:clres}),
consider the result from \citet{Tuccietal04} (their \S 3 and
Appendix).
\begin{eqnarray}
C_{\ell}^{res,fg,XY}\!\!\!\!\!\!&\!\!\!\!\!\!=\!\!\!\!\!\!&\frac{(\ln(\nu_0/\nu))^2}{16
\pi}\nonumber \\
&
\!\!\!\!\!\!\times& \!\!\!\!\!\!\sum_{\ell_1}(2\ell_1+1)C^{fg,XY}_{\ell_1}\sum_{\ell_2}(2\ell_2+1)C_{\ell_2}^{\alpha_{fg}}\left(^{\ell
\,\,\,\ell_1\,\,\ell_2}_{2 -2\,\,\, 0}\right)^2\nonumber \\
& +& C_{\ell}^{XY,\Delta
fg}\left(\frac{\nu}{\nu_0}\right)^{-2 \langle\alpha_{fg}\rangle}
\label{eq:fg1}
\end{eqnarray}
where $C_{\ell}^{\alpha_{fg}}$ denotes the power spectrum of the
spectral index error. If the power spectrum of the spectral index
error has the same $\ell$ dependence as the total intensity power
(that is $\sim \ell^{-3}$ for synchrotron and $\sim \ell^{-2.6}$ for
dust), neglecting the logarithmic dependence on $\nu$ we obtain
Eq. \ref{eq:clres}.
\begin{figure*}
\includegraphics[scale=0.55]{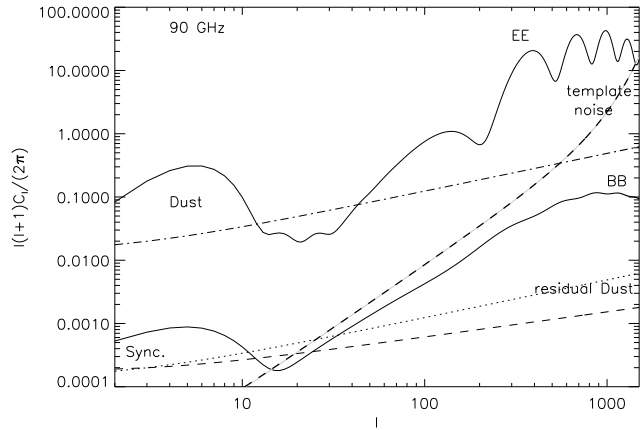}
\caption{ $EE$ and $BB$ signal for $r=0.03$ fiducial model, maximum estimated
  foreground level and foreground residuals for 1\% foreground subtraction at 90 GHz. Synchrotron is
  normalized to the DASI upper limit at $\ell=300$ $30$GHz assuming a
  frequency dependence of $\nu^{-3}$. Dust is normalized so that at 94 GHz,
  $\ell=900$ dust is 5\% polarized. The template noise is that expected for the  hypothetical satellite experiment.}  
\label{fig.fg}
\end{figure*}
We assume that the foreground template is computed for the
``cosmological'' frequency where the foreground contamination is
highest. For example, for 5 different frequency bands at ($30$, $50$,
$90$, $100$, $200$ GHz), the synchrotron template will be computed for
the $30$ GHz band and the dust template for the $200$ GHz band.

\subsection{Delensing}
Several different techniques to reconstruct the gravitational lensing
potential from CMB data and reduce or remove lensing contamination to
the $BB$ signal have been proposed
\citep{HuOkamoto02,KesdenCoorayKamionkowski02,KnoxSong02,
OkamotoHu03,KesdenCoorayKamionkowski03}.  However, in presence of
instrumental noise the contamination cannot be easily removed. In
addition foreground emission is expected to be highly non-Gaussian and
the performance of delensing techniques have not been explored in the
presence of non-Gaussian foregrounds. In fact, the separation of the
primordial gravitational wave contribution is achieved exploiting the
non-Gaussianity of the lensed CMB.  Thus, the non-Gaussian nature of
the foregrounds will degrade this reconstruction and limit the
delensing performance.  To include the above considerations, forecast
whether delensing can successfully be implemented and can improve
constraints on $r$ we proceed as follows. If foregrounds are
neglected, we assume that the delensing technique can be applied only
in the signal-dominated regime, and that in this regime the procedure
can reduce the BB-lensing signal only down to the level of the
instrumental noise power spectrum. This assumption is well motivated
by results of e.g., \citet{SeljakHirata04}. When including
foregrounds, we assume that delensing can be implemented only if the
$BB$ power spectrum of the combination of foreground emission and
foreground template noise is below 10\% of the $BB$ lensing power
spectrum, and that the $BB$-lensing signal can be reduced only down to
10\% of the foreground signal (emission plus template noise). We deem
this to be a reasonable assumption: a percent level of foreground
contamination in the angular power spectrum is probably close to the
minimum achievable and it corresponds to a 10\% level in the maps. The
residual foreground will be highly non-Gaussian thus limiting the
delensing to this level .  When considering a realistic case with
noise and foregrounds, for every $\ell$ we consider whichever effect
is most important.

\section{Cases we consider}\label{sec:experiments}
We consider several different experimental settings. First an ideal
experiment (IDEAL), then a more realistic full-sky satellite
experiment (SATEL) and finally,  ground-based and balloon-born partial sky experiments.
For all these cases, we compute the marginalized errors on the
parameters using the Fisher matrix approach illustrated above. In the
main text of the paper we report errors for the parameters relevant
for inflation and parameters degenerate with these (i.e. $r$, $n$,
$n_t$, $dn/d\ln k$ and $Z$).  The errors on the other cosmological
parameters are reported in the Appendix. For each case we start by
considering ten cosmological parameters ($r$, $n$, $n_t$, $dn/d\ln k$,
$Z$, $w_b$ $w_c$, $h$, $\Omega_k$ and $A$). We do not consider
calibration uncertainties, even though they propagate into the error
on the amplitude. As exploring non-flat models is computationally
expensive when we find that $\Omega_K$ is only degenerate with $h$, we
impose flatness (FLT). This constraint does not affect the estimated
error on the remaining cosmological parameters. Finally, we report errors
obtained with and without imposing the consistency relation ($n_t=-r/8$, CR).

\subsection{Ideal experiment}
Let us consider an ideal experiment, covering the full sky, with no
instrumental noise and no foregrounds. We consider $\ell_{max}$=1500
because smaller angular scales will be observed from the ground and,
most importantly, because higher multipoles in the temperature are
affected by secondary effects while in the polarization  $\ell >1500$
does not add significant cosmological information.

Since \citet{HirataSeljak03,SeljakHirata04} showed that the lensing
contribution to the $BB$ spectra can be greatly reduced and, in the
absence of instrumental noise and foregrounds, completely eliminated,
we consider two cases: the $BB$ spectrum with lensing (L) and without
(NL) (i.e. with the lensing contribution perfectly subtracted out).
                                   
We report the results in table Table~\ref{table:rfull}
In the L case, due to the lensing contamination, the same error can be achieved by
 considering only $\ell \leq 300$ for $BB$.  Errors on other relevant
 cosmological parameters are reported in Table \ref{table:rfull}.  We
 find that $r$ is mainly degenerate with $n_t$; by fixing $n_t$, the
 error on $r$ is greatly reduced (see Table \ref{table:rntfixed}), but
 the errors on the other cosmological parameters do not change
 significantly. 

If lensing can be completely subtracted from the $BB$ spectrum, we
obtain that the error bars on $r$ improve by more than an order of
magnitude (see Table~\ref{table:full}).

 \begin{table}
 \caption{\label{table:rfull} 1-$\sigma$ errors for an ideal
 experiment, including lensing (L), with no lensing (NL).}
 \begin{tabular}{ccccccc}
 &$r$ & $\Delta r$ & $\Delta n$ & $\Delta n_t$& $\Delta dn/d\ln k$ &$\Delta Z$\\
 \hline 
 &$0.01$&$0.001$ &  $0.0017$ &   $0.056$&     $0.003$ &    $0.003$ \\ 
L &$0.03$ &$0.0027$ &$0.0017$&   $0.047$ &     $0.0036$ &  $0.003$\\ 
 & $0.1$ &$0.006$ & $0.002$&   $0.035$ &      $0.0035$ &  $0.0035$ \\
\hline
 &$0.01$&   $0.000021$   & $0.0021$  &$0.0019 $ &$ 0.0038 $  &$ 0.0038$   \\ 
NL &$0.03$ &$0.000063$ &$0.0021$&   $0.0019$ &     $0.0038$ &  $0.0038$\\ 
\hline

\end{tabular}
 \end{table}

 \begin{table}
 \caption{\label{table:rntfixed} Ideal experiment, $\ell_{max}=1500$,
 $n_t$ fixed by consistency relation (CR)}
 \begin{tabular}{ccc}
$r$ & $\Delta r$ L & $\Delta r$ NL \\ 
\hline 
$0.01$  & $4.5\times 10^{-4}$ &  $1.6\times 10^{-5}$\\ 
$0.03$ &  $7.4\times 10^{-4}$ &  $4.9\times 10^{-5}$ \\ 
\hline
 \end{tabular}
 \end{table}

\subsection{Space-based experiment}

We next consider a full-sky (space-based) experiment with a beam of
 FWHM 8 arcminutes, 5 frequency channels ($N_{chan}=5$)  at 30, 50,
 70, 100, 200 GHz, and a sky coverage of 80\%.  Since the Galactic
 emission is highly polarized we consider that a realistic Galactic
 cut will exclude about 20\% of the sky.  We start by assuming a noise
 level of 2.2 $\mu K$ per beam per frequency channel. This noise level
 is achievable with the next generation space-based experiments, and is
 well-matched to measure the first peak in a possible $r=0.01$ gravity
 wave signal in the cosmic variance-dominated regime. We will then explore the
 effect of a higher (lower) noise level on the errors on $r$.  As
 before, our maximum multipole will be $\ell_{max}=1500$.  Finally we
 will explore how the constraints can be improved by implementing
 delensing.

We first consider an idealized case where there are no foregrounds. 
\begin{table}
 \caption{\label{table:sat1} Realistic satellite experiment, no
 foreground, no lensing subtraction, marginalized errors,
 $\ell_{max}=1500$. The error on the running is $\Delta dn/d \ln
k=0.0046$ in all these cases.}
 \begin{tabular}{cccccc}
case &$r$ & $\Delta r$ & $\Delta n$ & $\Delta n_t$  \\ 
\hline 
&     $0.01$ &  $0.003$ &$0.0023$  &$0.098$  \\ 
NO FG&$0.03$ &  $0.0048$ & $0.0023$ & $0.069$ \\ 
&     $0.1$  &  $0.01$ &$0.0023$ & $0.056$  \\
\hline
&     $0.01$ &  $0.0011$ &$0.0023$ & --  \\ 
NO FG&$0.03$ &  $0.0017$ &$0.0023$ & -- \\ 
CR&   $0.1$  &  $0.0028$ &$0.0023$ & -- \\
\hline
\end{tabular}
\end{table}
The constraints on $r$, $n$, and $n_t$ are reported in table
\ref{table:sat1}, the error on  $dn/d \ln k$ is $\Delta dn/d \ln
k=0.0046$ in all these cases.  When imposing flatness these constraints
are virtually unchanged.  Marginalized constraints on the other
parameters are reported in the Table \ref{table:full}.
 
Thus, we see that in the absence of foreground contamination, a
realistic experiment with this noise level can achieve constraints
close to those for an ideal experiment for our fiducial values of $r$.
We find that the noise level can be higher by a factor 10 before the
constraints get significantly degraded. In particular we find that for
$r=0.03$  a noise level of 22 $\mu$ K per beam  increases the error
on $r$ by an order of magnitude with respect to the zero-noise case.

Since the noise is not the limiting factor for the values of $r$
considered here, we conclude that reducing the sky coverage of such
an experiment to reduce the noise level would not improve the signal
to noise.

\subsubsection{Delensing}

To take into account of the possibility of improving the
signal-to-noise by applying the ``delensing'' (DL) technique we
proceed as follows. We start by ignoring foregrounds and consider only
the effect of instrumental noise.  Since in the noise-dominated regime
(i.e. when the power spectrum of the noise is greater than the
$BB$-lensing power spectrum), we assume that delensing cannot be
successfully implemented, we find\footnote{For greater computational
speed we have fixed $\Omega_k$ to be zero for this calculation. We
find that this affects only the error on the Hubble parameter which is
not the main focus of the present work.} that for our realistic
satellite experiment, delensing does not significantly improve the
constraints on the relevant parameters. In other words, for the noise
level considered, the signal-dominated regime for lensing is too small
to help improve constraints significantly.

Thus, in the absence of foregrounds, reducing the sky coverage to
lower the noise would help such a space-based experiment to access the
primordial signal at the ``peak''. However, partial sky experiments
can be done much more cheaply from the ground or from a balloon and
can reach even lower noise levels. We shall see below how these
conclusions can change in the presence of foregrounds.
  
\subsubsection{Foregrounds}

We find that even in the pessimistic case for the foreground amplitude
(for both dust and synchrotron), if foregrounds can be subtracted at
the 1\% level  ($\sigma^{fg,XY}=0.01$) and channels can be optimally
combined to minimize the effect of the residual contamination, the
additional noise-like component $C_{\ell}^{res}$ is comparable to or
smaller than the instrumental noise of 2.2 $\mu K$ per beam.  In
particular, we find that the total effective noise (that is,
instrumental noise plus foreground subtraction residuals) is higher
than the instrumental noise contribution by a factor $\sim 3$ at
$\ell=2$, but only a factor of $\sim 1.4$ at $\ell=10$ and $\sim 1.1$
at $\ell>100$.   As a result, the constraints on cosmological
parameters are not significantly affected by the presence of a
residual foreground contamination at this level.  Note that we do not
propagate into the error bars a possible uncertainty on the residual
foreground contamination, as this will be a higher-order effect in our
approach; in other words we assume that the uncertainty in the
residual foreground contamination is negligible compared with the
residual contamination level. When dealing with realistic data the
validity of this assumption will need to be assessed.

In order for the foregrounds to be subtracted at the percent level, a
drastically improved knowledge of the amplitude and the spectral and
spatial dependence of polarized foreground emission is needed. This
can be achieved by observing the polarized foreground  emission  on
large areas of the sky, at high galactic latitudes, with high
resolution and high sensitivity and at multiple wavelengths, both at
higher and lower frequencies than the ``CMB window''.  Most likely, a
combination of approaches will be needed.  For example, the Planck
satellite will observe the full sky polarization up to 350 GHz and
down to 30 GHz. The proposed BLASTPol (M. Devlin, private
communication)  will survey smaller regions of the sky at higher
sensitivity and at higher frequencies, enabling one to extrapolate the
dust emission more securely.  The ability to create accurate
foreground templates and to reduce foreground contamination will
depend crucially on the success of these efforts.

As a foreground residual of 1\% may be too optimistic, we also consider
cases with a 10\% foreground residual in the power spectrum.  We 
find that,  while  for level of foreground residuals of 10
constraints on the recovered parameters degrade rapidly.

Our fiducial values for $r$ ($0.01$, $0.03$ and $0.1$) have been
chosen because  they are accessible to experiments in the very near
future.  A detection of such a large primordial tensor component would
support a large-field model (see also the discussion of hybrid
inflation models in \S\ref{sec:theory}). Following the considerations
of \citet{efstathiou/mack:2005}, it would imply that, if the inflaton
is a fundamental field, the effective
field theory description is not valid, i.e. that  Eq. \ref{eq:effptl}
is not a self-consistent description of inflation.

Thus, we also investigate lower values of $r$ of $10^{-3}$ and
$10^{-4}$ which are closer to satisfying the effective field theory
description (see Table~\ref{table:satlowr}).  We find that a
three-$\sigma$ detection of $r=10^{-3}$ is realistically achievable if
the foreground contamination is lower than the DASI upper limit and if
foreground cleaning can reduce contamination at least to  the 1\%
level in the power spectrum.  However a value of $r=10^{-4}$ cannot be
detected by this realistic experiment given our noise level and
estimates of foregrounds.  To assess whether the limiting factor for
detecting such a gravity wave signal is the noise level or foreground
contamination, we computed the expected errors for a case with no
noise, synchrotron amplitude 1/2 of the DASI limit, dust polarization
fraction of 5\% and foreground subtraction at the 1\% level. Imposing
the consistency relation we find that $\Delta r=6\times10^{-5}$, not
even a two-$\sigma$ detection. This implies that the main obstacle to
detecting such a small value of $r$ will come from Galactic
foregrounds.  To assess quantitatively how much the $r$ detection
could be degraded by a lower value of $\tau$ we also report the errors
for a fiducial model with $r=0.01$ and $\tau=0.1$.

As the delensing implementation in this case is limited by foregrounds
residuals, not by the noise level, it would not help for a space-based
experiment to reduce the sky coverage by a factor of a few, say, to
reduce the noise levels. As we will see below, delensing could be
implemented by targeting a small region of the sky with particularly
clean foregrounds. However, this can be done more easily and cheaply
from the ground.

Finally we note that, although we used the full combination of $T$, $E$
and $B$ data, the signal for $r$ comes from the $BB$ signal. In fact, we
find that if we consider our realistic satellite case but only $T$ and
$E$-mode polarization data, only upper limits can be imposed on $r$ for
$r<0.1$.

\begin{table}
\caption{\label{table:satlowr} Realistic satellite experiment
  including foregrounds for $r\leq 10^{-3}$ and flat cosmology. We
  have assumed that synchrotron emission amplitude is 50\% of DASI
  limit, a dust polarization fraction 5\% and that foregrounds can be
  subtracted to 1\% in the power spectrum. For comparison we also
  report a case for no instrumental noise. In the latter case, the
  error on $r$ is reduced only by a factor $\sim 2$, indicating that
  foregrounds contamination becomes the limiting factor.}
\begin{tabular}{c|ccccc}
 case& $r$& $\Delta r$ & $\Delta n$ & $\Delta n_t$& $\Delta dn/d\ln
 k$\\
\hline
SAT &$0.0001$&$0.00019$&$0.0023$&$0.43$&$0.0046$\\
    &$0.001$&$0.0013$&$ 0.0023$&$0.31$&$0.0046$\\
\hline
+CR &$0.0001$&$0.00012$&$ 0.0023$&--&$0.0046$\\
    &$0.001$&$0.0003$&$ 0.0023$&$-$&$0.0046$\\
\hline
SAT $\tau=0.1$&$0.001$&$0.0020$&$0.002$&$0.47$&$0.0045$\\
+CR&          $0.001$&$0.00043$&$0.002$&--&$0.0045$\\
\hline
\hline
NO NOISE& $0.0001$&$0.00007$&$0.0019$&$0.17$&$0.0037$\\
+CR & $0.0001$&$0.000057$&$0.0019$&--& $0.0037$\\
\hline
\end{tabular}
\end{table}

\subsection{Ground-based and balloon-borne experiments} 

While a space-based experiment may be a decade away, in the shorter
term, partial-sky ground-based experiments will be operational in the
next few years.  Ground-based experiments are less expensive and can
achieve a lower level of noise contamination at the expense of
covering a smaller region of the sky. However, as the primordial
signal is minimal at $10<\ell<30$, the signal-to-noise is expected to
scale only as the square root of the sky fraction for experiments
covering more that about 200 square degrees (for the same noise level,
frequency coverage and angular resolution).

For $r=0.03$, the primordial $B$-mode signal dominates the
lensing one on scales of 5-10 degrees ($\ell \lap 70$), but for
$r=0.01$, it dominates only on scales larger than  few $\times$ 10
degrees ($\ell \lap 10$).  So ground-based experiments may need to
target a clean patch of the sky and will need to rely on delensing to
detect $r$ values $\lap 0.03$.

On the other hand, the amplitude of the large-scale $B$-mode signal
strongly depends on the value for $\tau$; thus a limit on $r$ from
these scales is somewhat more cosmology-dependent.

Here we consider several ground-based and balloon-borne experiments;
the full set of experimental specifications assumed for each of these
are reported in the Appendix (Table~\ref{table:experiments}). 

Continuously changing the sky coverage, frequency coverage and number
of frequencies, number of detectors and angular resolution for these
experiments will be prohibitive. But we want to concentrate on
experimental setups that can be made with next generation technology;
hence we report here results for a few selected experimental designs
that we found already make the best use of next generation
technology. We will then consider small variations around these
setups. We will also discuss in the text results for selected
combinations of $r$ and foreground assumptions and report all other
cases considered in Table $6$.

Firstly, we consider two instruments with the specifications of the
ground-based experiments QUIET and PolarBeaR (based on HEMPT and
bolometer technology respectively), and two possible combinations of
these experiments, ensuring wider frequency coverage. Of these,
QUIET+PolarBeaR is a straight combination of the frequency channels
and noise properties of QUIET and PolarBeaR. QUIETBeaR is a
hypothetical experiment with the detectors of QUIET with added
channels at the PolarBeaR frequencies, but with the QUIET 90 GHz noise
properties, designed to observe a significantly smaller area of the
sky. The motivation behind this combination is to check the
improvement in constraints in an experiment where it is possible to
successfully implement delensing.  Finally, we consider a
balloon-based experiment with the properties of SPIDER (J. Ruhl
private comm.).

The ground-based experiments considered above obtain their $r$
constraint principally by going after the primordial $BB$ ``first
peak'' in the $\ell$--range 70-100. On the other hand, the larger sky
coverage of the balloon-borne experiment makes it possible for it to
access the ``reionization bump'' at large scales, where the primordial
signal may dominate significantly over the weak lensing, depending on
the value of $\tau$. Therefore, this selection of experiments pretty much cover
the possibilities of what can be done without going to space.

We find that QUIET can detect $r=0.1$ at the $3$-$\sigma$ level when
imposing the consistency relation, if foregrounds can be subtracted at
the 1\% level (see Table~\ref{table:grnd}). Alternatively if
foregrounds can only be cleaned at the 10\%, the experiment should
observe a clean patch of the sky (with minimal foreground emission)
needs to be observed to achieve the same signal-to-noise.

Since we assume that delensing can be implemented only where the
$B$-lensing power spectrum is 10 times higher than the foreground
emission and template noise, we find that delensing cannot be
implemented for our pessimistic foreground levels. It could be
implemented for a clean patch of the sky where the synchrotron
amplitude is 1/10 of the DASI limit and dust polarization fraction is
$\sim$ 1\%.  In this case, the limiting factor for delensing becomes
the noise in the foreground templates. We find that the error-bars 
are not significantly changed and $r=0.01$ is still out
of reach (Table~\ref{table:full}).

An experiment like PolarBeaR can improve on these constraints on $r$
if the dust contamination is minimal (1\% of dust contamination according to
our assumptions; Table~\ref{table:grnd}).

However, an aggressive level of foreground cleaning, as the one considered 
here, can realistically be achieved only with a wider frequency coverage
well below and above the cosmological window.  Moreover, additional
frequencies enable one to reduce the foreground template noise and
make delensing possible. The QUIETBeaR  and QUIET+PolarBeaR channel
combination, with the addition of higher frequencies, improves dust
cleaning; lower frequencies will be useful for improving  synchrotron
subtraction. 
We find that the QUIET+PolarBeaR combination in a clean patch of sky
(50\% of DASI limit and 1\% dust polarization) with good foreground
subtraction can detect $r=0.1$ at better than the 3-$\sigma$ level and
$r=0.03$ at about 3-$\sigma$ level.

On the other hand, delensing can be implemented if a small, clean
patch of the sky is observed with high sensitivity.  In our analysis,
if an experiment like QUIETBeaR can achieve 1\% foreground cleaning
and can observe a particularly clean patch of the sky, we forecast
that delensing can be successfully implemented; thus $r=0.01$ can be
detected at the $\sim 3$-$\sigma$ level if imposing the consistency
relation (Table~\ref{table:grnd}).  

Note, however, that at small angular scales, extragalactic point
source contamination -- which we have neglected -- may have an
amplitude comparable to that of the lensing signal, making delensing
implementation even harder.

\begin{table}
\caption{\label{table:grnd} Ground-based and balloon-borne experiments
  (see Table~\ref{table:experiments} for specifications), including
  instrumental noise and foregrounds, for flat
  cosmologies. $\tau=0.164$ has been taken to be the fiducial value
  unless explicitly stated otherwise. We have used WMAP priors on
  cosmological parameters as explained in the text.}
\begin{tabular}{c|ccc}
 case& $r$& $\Delta r$ & $\Delta n$ \\
\hline
QUIET &$0.01$&$0.015$&$0.029$\\
FG 1\%&$0.03$&$0.019$&$0.03$\\
CR    &$0.1$ &$0.035$ &$0.03$\\
\hline
QUIETBeaR& &&\\
DASI10\%,dust1\%&$0.01$&$0.003$&$0.054$\\
FG1\% CR DL&&&\\
\hline
PolarBeaR&      $0.01$&$0.012$&$0.018$\\
DASI50\%,dust1\%&$0.03$&$0.015$&$0.018$\\
FG1\% CR &      $0.1$ &$0.023$&$0.018$\\
\hline
QUIET+PolarBeaR& $0.01$&$0.006$&$0.015$\\
DASI50\%,Pol1\%&$0.03$&$0.009$&$0.02$\\
FG1\% CR &$0.1$&$0.017$&$0.02$\\
\hline
SPIDER, $\tau=0.1$&&&\\
DASI50\%,Pol1\%&$0.01$&$0.004$&$0.1$\\
FG1\% CR &&&\\
\hline
\end{tabular}
\end{table}

The specifications of the SPIDER experiment are not too dissimilar
from those of the realistic satellite experiment considered above. The
slightly different frequency coverage and noise level, the different
beam size and the fact that the remaining fraction of the sky after
the Galactic cut is 40\%, change the forecast errors only
slightly. For example for a fiducial model with $\tau=0.1$, $r=0.01$
can be detected at the 2-$\sigma$ level in the presence of noise and
foregrounds.  Thus, the same considerations for the satellite
experiments apply here. We report forecasts for SPIDER in
Tables~\ref{table:grnd} and \ref{table:full}. Note that the errors on
other parameters which are mostly constrained by the temperature data
at $\ell >300$ such as $n$, $w_b$ etc. are larger than for our
satellite experiment. This is because the beam size of SPIDER is not
optimized for this. The error on $r$ for the CR case is only a factor
of a few larger than for a satellite experiment for $\tau=0.1$. Such
an experiment could perform two flights, one in the northern
hemisphere and one in the southern hemisphere, and thus cover a larger
fraction of the sky. Assuming that the two maps can be accurately
cross-calibrated we find that the errors on $r$ could be reduced
scaling approximately like $\sqrt{f_{sky}}$.

\section{Conclusions}\label{sec:conclusions}

One issue that we have not discussed is that, while on the full sky
$E$ and $B$ modes are separable, this is no longer the case when only
small patches of the sky are analyzed, as the boundaries of the patch
generate mode mixing. The smaller the patch, the more important the
effect of the boundary is expected to be.  In the absence of
foregrounds,
\citet{LewisChallinorTurok02} showed that for simple patch geometries,
$E$ and $B$ modes can still be separated. In particular, for circular
patches larger than about 5 degrees in radius, they find that the
degradation due to imperfect $E$--$B$ separation is
negligible. However, for more general patch shapes this may be
optimistic. All the experiments we considered had patches comfortably
larger than this limit, and therefore we expect that this effects
should not degrade our forecasts significantly.

From the above results we can deduce some general considerations which
may help in planning, designing and optimizing future $B$-mode
polarization experiments.  Given the realistically achievable
constraints on $r$, we can then forecast what can be learned about
inflationary physics from these experiments. In our forecast, we have
also neglected ``real world'' effects that are experiment-specific
such as inhomogeneous/correlated noise, $1/f$ effects, etc. But these
effects are not expected to significantly alter our conclusions
(POLARBeaR and QUIET collaborations, private comm.).

\subsection{Guidance for $B$-mode polarization experiments} 

Our assumptions about foregrounds contamination are based on
information coming from much lower or higher frequencies than the CMB
window, and in many cases from observations of patches of the
sky. Since the properties of the polarized foregrounds show large
spatial variations across the sky, the optimal ``cosmological window''
may depend strongly on the area of the sky observed and may be
different for full sky or partial sky experiments.  In addition,
experiments working below $\sim70$ GHz will want to concentrate in sky
regions with low synchrotron contamination, while experiments at
higher frequencies need to focus regions with low dust: one ``recipe''
may not ``fit all''.  Due to our limited polarized foreground
knowledge it is not yet possible to make reliable predictions on which
areas of the sky will be the cleanest and more suitable for primordial
B-mode detection (or if the experiments considered here, by reducing
their sky coverage, could be less contaminated by
foregrounds). However forthcoming datasets (e.g., WMAP polarized maps)
combined with existing ones (e.g., Archeops and Boomerang polarization
maps) will enable one to do so.
 
The primordial $B$-mode signal on large scales (accessible by
space-based or balloon-borne experiments) is not contaminated by
lensing, but its amplitude is affected by the optical depth to
reionization $\tau$.  The lensing signal (dominant on scales $\ell\gap
50$) can be removed to high accuracy (delensing) in the absence of
noise and foregrounds.  Foreground contamination and noise in the
foreground templates are the limiting factor in constraining $r$ and
in the delensing implementation. For the considered realistic
space-based experiments the residual noise (both in the maps and in
foreground templates) hinders delensing implementation. Delensing may
be used to improve $r$ limits in partial sky ground-based experiments:
partial sky experiments can easily achieve lower noise level than
space-based ones and can target particularly clean areas of the sky,
but accurate foreground templates are still needed to keep foregrounds
contamination at or below the 1\% level and to reduce template noise.

A space-based or balloon-borne experiment can easily measure
$r=10^{-3}$ (at the $\sim$3$-sigma$ level) if foregrounds can be
subtracted at 1\% level, but foreground contamination is ultimately
the limiting factor for a detection of $r\sim 10^{-4}$.

From the ground, $r=0.01$ can be detected at the 2-$\sigma$ level but
only if foreground can be subtracted at the 1\% level and if accurate
foreground templates are available.

In order to have accurate and high resolution foreground templates and
for the foregrounds to be subtracted at the percent level, a
drastically improved knowledge of the amplitude and the spectral and
spatial dependence of polarized foreground emission is needed. This
can be achieved by observing the polarized foreground emission on
large areas of the sky, at high Galactic latitudes, with high
resolution and high sensitivity at multiple wavelengths, both at
higher and lower frequencies than the ``CMB window''.  The success of
future B-modes experiments will rely on these efforts.  On-going and
planned experiments such as EBEX, CLOVER, QUaD, Planck, BLASTPol, will
play a crucial role in achieving this goal.

If this can be achieved, in order to constrain $r$ as best as
possible, a combination of approaches will most likely be needed. A
space-based or balloon-borne experiment can be  optimized to observe
the low $\ell$ $BB$ ``bump'': it can have a wide frequency coverage,
but weight restrictions can be met by having less stringent
requirements for noise and angular resolution.  Future advances to
ballooning methods, such as ``stratellite''\footnote{Stratellite: {\tt
http://www.sanswire.com}} technology may make that avenue particularly
attractive. A stratellite has a flight time of 18 months' duration, is
stationed at 65,000 feet and is capable of carrying a  payload of up
to 3000 lb. In addition the airship is 100\% reclaimable and the
vehicle will be much cheaper to build and to run than a satellite.
These properties may make such an experiment achieve the sky coverage
and integration time of a satellite, retaining at the same time  the
advantages of a balloon-borne experiment (e.g. lower costs, less
weight restrictions, upgradable).

A satellite/balloon-borne experiment is nicely complementary to a
ground-based partial-sky experiment.  The ground-based telescope can
be optimized to implement delensing: it can observe a particularly
clean patch of the sky; and can achieve low noise level by using large
detectors arrays. Thus the noise and beam size requirements need to be
more stringent than for a full sky experiment tailored to access the
reionization bump, and the frequency coverage must be wide enough to
still enable accurate (percent level) foreground subtraction.

\subsection{Implications for Inflation}\label{sec:conclusions:inf}
\begin{figure}
\includegraphics[scale=0.35]{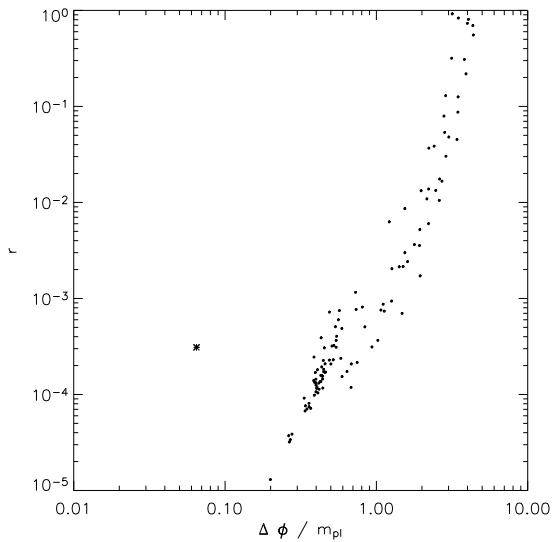}
\caption{Otherwise identical to Fig. \ref{fig.lythboundwmap} except that the
  constraints on $n_s$ and $d n_s / d \ln k$ are 2-$\sigma$ limits from the CR
  case of Table \ref{table:sat1}.} 
\label{fig.lythboundwmapspace}
\end{figure}

\begin{figure}
\includegraphics[scale=0.35]{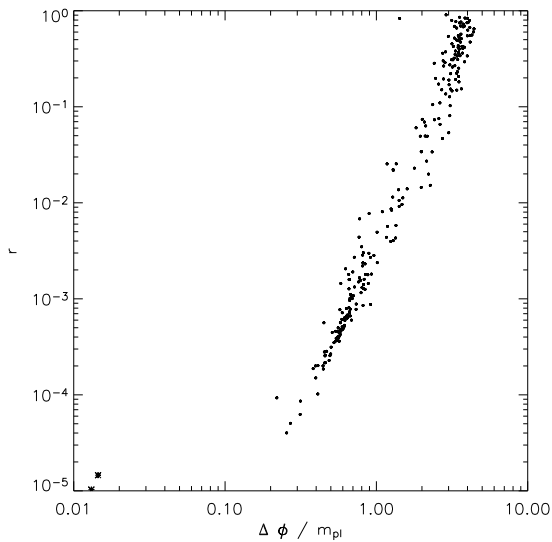}
\caption{Otherwise identical to Fig. \ref{fig.lythboundalldata} except that the
  constraints on $n_s$ and $d n_s / d \ln k$ are 2-$\sigma$ limits from the CR
  case of Table \ref{table:sat1}.} 
\label{fig.lythboundalldataspace}
\end{figure}

What can experiments with the capabilities illustrated here tell us
about inflationary physics? Figures \ref{fig.lythboundwmap} and
\ref{fig.lythboundalldata} show that, to produce a clearly detectable
($>$3-$\sigma$) tensor component in any foreseeable CMB experiment,
inflation must necessarily involve large-field variations: $\Delta\phi
\gap 1$. As \citet{efstathiou/mack:2005} point out, the relation in
Eq.~\ref{eq:lythflow} is so steep that, to probe models with small
field variations where an effective field theory description is likely
to be valid, one needs to be able to detect $r \le 10^{-4}$. This is
exceedingly difficult in CMB experiments, given realistic polarized
foregrounds, detector noise achievable in the foreseeable future, and
the weak lensing contamination of the primordial $B$-modes. Therefore,
in the foreseeable future, it appears that we are limited to testing
large-field models of inflation. Figures \ref{fig.lythboundwmapspace}
and \ref{fig.lythboundalldataspace} show what happens to the
$\Delta\phi$ vs. $r$ relation when the observational limits on $n_s$
and $d n_s / d \ln k$ at the level of the 2-$\sigma$ limits from the
CR case of Table \ref{table:sat1} are imposed.
Observation of a significant tensor/scalar ratio may therefore be a
signal that inflation is being driven by some  physics in
which the inflaton cannot be described as a fundamental field captured
in terms of a low energy effective field theory description such as Eq.~\ref{eq:effptl}.
Note that the converse of this argument is not implied; there are
perfectly valid inflation models realized within the effective field
theory framework in which the inflaton is not a fundamental field. Alternatively, as \citet{Linde:2004kg} points out, the
effective field theory description of $V(\phi)$ becomes invalid only
for $V(\phi) > m_{\rm Pl}^4$, not $\phi > m_{\rm Pl}$ as inferred from
Eq. \ref{eq:effptl}. Chaotic inflation models, such as the simple
$\lambda \phi^4$ model (excluded at about the 3-$\sigma$ level by
current cosmological constraints), exploit this rationale; on the
other hand, building realistic particle-physics motivated chaotic
inflation models has historically proved difficult. However, it has
been shown recently that chaotic inflation models can be realized in
supergravity models if the potential has a shift symmetry (see
\citet{Linde:2004kg} and references therein for a detailed
explanation). Thus it might be possible to construct chaotic inflation
models motivated by realistic particle physics with $\Delta\phi \gap
1$. 
Further, as \citet{Boyanovskyetal05} point out, inflation can be described
 as a series expansion in terms of a new variable
 $\chi=\phi/(\sqrt{N}m_{\rm Pl})$, which makes the effective field
 theory description valid up to $r\sim 1$. In this case $\phi$ is not
 a fundamental field but a degree of freedom in the theory.

It is therefore important to investigate the possibility of
constructing particle-physics motivated inflationary models with
$\Delta\phi \gap 1$, since these are likely to be the only models that
could be probed by realistic CMB experiments in the near
future. Significant progress in this direction has already been made;
for example, extranatural/pseudonatural inflation \cite{Hill:2002kq,
Arkani-Hamed:2003wu, Arkani-Hamed:2003mz}, in which the inflaton is
the extra component of a gauge field in a 5-dimensional theory
compactified in a circle, which predicts a significant production of a
tensor component ($r \sim 0.11$). Purely 4d theories appear to require
more sophisticated structures in order to protect the flatness of the
potential from excessive radiative corrections \citep{Kim:2004rp,
Arkani-Hamed:2003mz}, and in general do not predict significant
gravity-wave production. However they still appear to preserve the
prediction of a significant deviation from scale invariance as a
signature of inflation.

Conversely, if tensor modes are not detected in these experiments, we
can exclude large-field models as a mechanism for inflation. Then we
would need to examine the remaining possibilities by investigating
constraints on the deviation from scale-invariance of the scalar
spectral index (a blue or red index discriminates between classes of
inflationary models), and also from its running. While this paper
focuses on the implications of a primordial $B$-mode measurement from
foreseeable CMB experiments for inflation, a negligible primordial
tensor component is also a prediction of the Ekpyrotic/Cyclic
scenarios \citep{Khoury:2001bz, Khoury:2001wf}.

\begin{figure}
\includegraphics[scale=0.5]{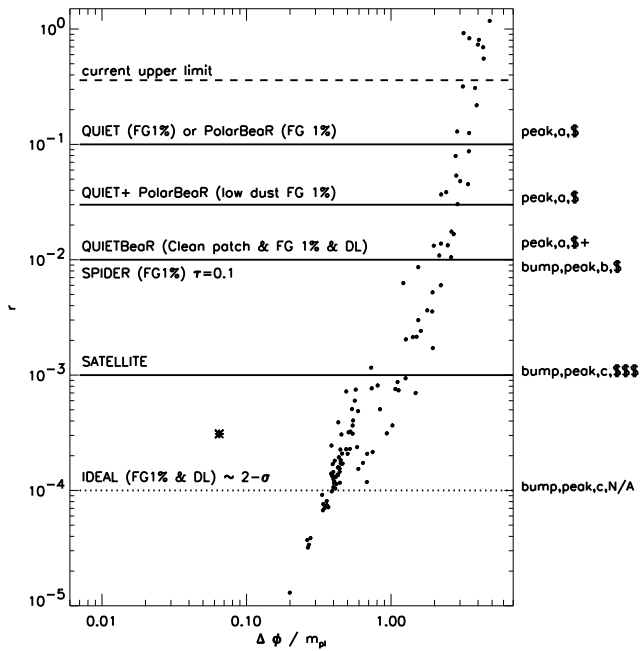} 
\caption{Superimposed upon the Monte-Carlo points of
Fig.~\ref{fig.lythboundwmapspace} are the minimum values of $r$ that
can be measured at the $3$--$\sigma$ level for different experimental
setups and foreground knowledge. The legend on the right indicates
whether the signal comes from the ``peak'' ($\ell >30$) region or the
``bump'' ($\ell <30$) region; if the experiment is ground-based (a),
balloon-borne (b) or satellite (c); and how expensive such an
experiment may be.  The top dashed line shows the current upper
limit. Single ground-based experiments can reach $r=0.1$ if foreground
subtraction can be done at the 1\% level. The signal in this case
would come from the peak region and the noise characteristic would not
allow delensing. If two different experiments such as QUIET and
PolarBeaR can {\it i)} overlap their survey region and {\it ii)}select
a sky patch with realistic synchrotron contamination (50\% of DASI
upper limit) and low polarized dust emission (1\% of Archeops limit)
and {\it iii)} clean foregrounds to the 1\% level, the wider frequency
coverage and the lower noise means that $r=0.03$ can be reached. Also,
in this case the signal comes from the peak region.  $r=0.01$ can be
reached in two different ways. The technologies of QUIET and PolarBeaR
could be combined: QUIET noise properties can be extended to also
include PolarBeaR channels, a smaller region of the sky is scanned so
that the noise per pixel can be reduced and delensing can be
implemented; and the sky region chosen is particularly clean (dust:
1\% of Archeops limit; synchrotron: 10\% of DASI limit). The signal
would come from the ``peak'' and the measurement would rely on
delensing. Alternatively a long duration balloon flight with the
specifications of SPIDER would measure the same $r$ but accessing the
reionization bump region and without relying on delensing. $r=10^{-3}$
can be achieved only with a satellite mission. Finally, foreground
contamination, even if at the 1\% level, is the limiting factor for
accessing $r= 10^{-4}$ with an ideal, space based, noiseless
experiment.}
\label{fig:bangforbucks}
\end{figure}

\subsection{``Bang for your bucks''}
To conclude, in Figure \ref{fig:bangforbucks} we summarize the minimum
experimental set up and foreground knowledge that is needed to measure
a given $r$ value at $\gap 3$--$\sigma$ level. We also indicate
whether the signal comes from the reionization ``bump'' region
(affected by $\tau$) or the ``first peak'' region (affected by lensing
contamination) and if delensing can be used to improve signal to
noise. Finally we label the experiments with letters {\it a} for
ground based, {\it b} for balloon-borne and {\it c} for
satellite. These could also be interpreted as price-tags from least
expensive (a) to most expensive (c).  There is a clear correlation
between cost of the experiment and models that can be accessed by it.

\section*{acknowledgments}
We thank Mark Devlin, George Efstathiou, Chao-Lin Kuo, Stephan Meyer,
Lyman Page, John Ruhl, Dorothea Samtleben, Kendrick Smith, David
Spergel, Huan Tran, Ben Wandelt, and Bruce Winstein for useful
conversations, and Eiichiro Komatsu and Adrian Lee for comments on an
early version of the manuscript. LV is supported by NASA grant
ADP03-0000-0092 and ADP04-0000-0093. HVP is supported by NASA through
Hubble Fellowship grant \#HF-01177.01-A awarded by the Space Telescope
Science Institute, which is operated by the Association of
Universities for Research in Astronomy, Inc., for NASA, under contract
NAS 5-26555. She acknowledges the hospitality of the Kavli Institute
for Particle Astrophysics and Cosmology at Stanford University, where
part of this work was carried out. The research of RJ is partially
supported by NSF grant AST-0206031.


\section*{Appendix}

\begin{table*}
\caption{\label{table:full}Constraints on all parameters. IDEAL
  denotes an ideal experiment(no foregrounds, cosmic variance
  dominated). SAT denotes a space-based experiment. CR means that the
  consistency relation for $n_T$, $n_t=-r/8$ has been imposed. FLT
  means that flatness has been imposed, this does not affect the
  errors on the parameters relevant to inflation. FG denotes that
  galactic foregrounds has been considered.  FG1\% means that their
  residual contamination is reduced to 1\% while FG10\% means that
  foreground contamination has been reduced to 10\%. The experimental
  specifications used for obtaining these constraints are reported in
  Table~\ref{table:experiments}.  The error on the amplitude does not
  include possible calibration errors which are likely to
  dominate. These forecasts have been computed for a fiducial value of
  $\tau=0.164$ except where explicitly specified. As the $BB$ signal
  on large scales is boosted by $\tau$ roughly as $\tau^2$, lower
  values of $\tau$ will degrade the significance of these
  forecasts. For fiducial  $\tau$ one sigma below the best fit value
  the detectability of $r=0.01$ ($r=0.03$) from $\ell < 20$ is
  equivalent to the detectability of $r=0.03$ ($r=0.1$) in the best
  fit model. This consideration affects only large sky coverage
  experiments which probe the ``reionization bump'' in the $BB$
  spectrum at $\ell <10$ and does not affect the smaller scale
  ground-based experiments.} 
\begin{tabular}{cccccccccccc}
$r$& case & $\Delta r$ & $\Delta n$ & $\Delta n_t$& $\Delta dn/d\ln k$ &$\Delta Z$&$\Delta \omega_b$ &$\Delta \omega_c$ &$\Delta h$ &
  $\Delta \Omega_K$& $\Delta_A$ \\ \hline \hline 
$0.01$  & &$0.001$ &$0.0017$ & $0.056$&$0.0034$ &$0.0033$ &$6\times 10^{-5}$ &$0.00027$  &$0.003$&$0.0006$ &$0.004$ \\ 
$0.03$ & IDEAL &  $0.0028$ & $0.0017$& $0.047$  &  $0.0036$ &  $0.0034$&$6.4\times 10^{-5}$ & $0.00026$&$0.0033$& $0.0006$&$0.004$ \\
  $0.1$  & &$0.0063$ &$0.0018$ & $0.035$&$0.0035$ &$0.0036$ &$6\times 10^{-5}$ &$0.00020$  &$0.0023$&$0.0006$ &$0.004$ \\ \hline
$0.01$  & & $0.00045$ & $0.0017$& -- & $0.0036$ &  $0.0032$&$6.1\times 10^{-5}$ & $0.00025$&$0.003$& $0.0006$&$0.0038$\\ 
$0.03$ & IDEAL,& $0.00074$ & $0.0017$ & -- &  $0.0036$ &  $0.0032$ &$6.4\times 10^{-5}$ & $0.00026$&$0.0032$ & $0.00060$&$0.0038$ \\ 
$0.1$  & CR &$0.0015$ & $0.0017$& -- &  $0.0036$& $0.0032$&$6.4\times 10^{-5}$ & $0.00026$&$0.0024$& $0.0006$&$0.0038$
   \\ \hline
$0.01$  & IDEAL& $0.000021$ & $0.0021$& $0.0019$ & $0.0038$ &  $0.0038$&$6.5\times 10^{-5}$ & $0.00072$&$0.0057$& $0.0006$&$0.0048$\\ 
$0.03$  &DL&$0.000063$ &$0.0021$&$0.0019$& $0.0038$ & $0.0038$&$6.5\times 10^{-5}$ & $0.00072$&$0.0057$&$0.0006$&$0.0049$\\
\hline
$0.01$  & IDEAL& $0.000016$ & $0.0021$& -- & $0.0038$ &  $0.0038$&$6.5\times 10^{-5}$ & $0.00072$&$0.0057$& $0.0006$&$0.0048$\\ 
$0.03$  &DL CR&$0.000049$ &$0.0021$&--& $0.0038$ & $0.0038$&$6.4\times 10^{-5}$ & $0.00072$&$0.0057$&$0.0006$&$0.0049$\\
\hline \hline
$0.01$  &  &$0.0030$ & $0.0023$& $0.098$ &  $0.0046$ &  $0.0053$&$8.4\times
   10^{-5}$ & $0.00053$&$0.0055$& $0.00078$&$0.0066$  \\ 
$0.03$ & SAT & $0.0048$&$0.0023$& $0.069$ &  $0.0046$ & $0.0053$&$8.4\times 10^{-5}$ & $0.00054$&$0.0056$& $0.00080$&$0.0066$ \\
$0.1$ &  & $0.010$ & $0.0023$& $0.056$ &  $0.0046$ &  $0.0055$&$8.4\times
   10^{-5}$ & $0.00055$&$0.0058$& $0.00082$&$0.0068$ \\
\hline
$0.01$  & SAT &$0.0030$ & $0.0023$& $0.096$ &  $0.0046$ &  $0.0049$&$8.3\times
   10^{-5}$ & $0.00050$&$0.0023$& --&$0.0062$  \\ 
$0.03$ &  & $0.0047$
   & $0.0023$& $0.068$ &  $0.0046$ &
   $0.0050$&$8.3\times 10^{-5}$ & $0.00051$&$0.0023$& --&$0.0062$\\
$0.1$ & FLT& $0.010$ & $0.0023$& $0.054$ &  $0.0046$ &  $0.0050$&$8.3\times
   10^{-5}$ & $0.00052$&$0.0023$& --&$0.0062$ \\
\hline
$0.01$  & & $0.0011$ & $0.0023$& -- &  $0.0046$ &  $0.0052$&$8.3\times
   10^{-5}$ & $0.00051$&$0.0054$&$0.00078$&$0.0065$\\ 
$0.03$ &SAT& $0.0017$ & $0.0023$&
   -- &  $0.0046$ &  $0.0050$&$8.3\times 10^{-5}$
   & $0.00051$&$0.0054$&$0.00078$&$0.0063$ \\
$0.1$  &CR& $0.0028$ & $0.0023$& -- &
   $0.0046$ &  $0.0049$&$8.2\times 10^{-5}$ &
   $0.00051$&$0.0054$&$0.00080$&$0.0062$ \\
\hline \hline
$0.01$ &SAT & $0.0031$& $0.0023$&
   $0.098$&$0.0046$ &$0.0049$&$8.3\times
   10^{-5}$&$0.0005$ &$0.0024$ &--& $0.0063$\\ 
$$ &FG1\% FLT & &&&&&&&&&\\ 

    $0.01$ &+CR &$0.001$ & $0.0022$& --
   &  $0.0046$ &  $0.0049$&$8.3\times 10^{-5}$ &
   $0.0005$&$0.0023$& --&$0.0062$\\
\hline
   $ $ &SAT, &&&&&&&&&&\\ 
   $0.01$ & DASI50\%, $\tau=0.1$ &$0.0030$ & $0.0021$& $0.1$
   &  $0.0044$ &  $0.0040$&$7.6\times 10^{-5}$ &
   $0.00047$&$0.0021$& --&$0.0045$\\
   $ $ &FG1\% FLT&&&&&&&&&&\\ 
   $0.01$& +CR&$0.0012$&$0.0020$&--&$0.0043$&$0.040$&$7.6\times10^{-5}$&$0.00046$&$0.0020$&--&$0.0045$\\ 
\hline
$0.0001$&SAT FLT&$0.00019$&$0.0023$&$0.43$&$0.0046$&$0.005$&$8.3\times10^{-5}$&$0.00051$&$0.0023$&--&$0.0063$\\
$0.001$&FG1\% DASI50\%&$0.0013$&$ 0.0023$&$0.31$&$0.0046$&$0.005$&$8.3\times10^{-5}$&$0.0005$&$0.0023$&--&$0.0063$\\
\hline
%
$0.0001$&SAT FLT CR&$0.00012$&$ 0.0023$&--&$0.0046$&$0.005$&$8.3\times10^{-5}$&$0.0005$&$0.0023$&--&$0.0063$\\
$0.001$&FG1\% DASI50\%&$0.0003$&$ 0.0023$&--&$0.0046$&$0.005$&$8.3\times10^{-5}$&$0.0005$&$0.0023$&--&$0.0063$\\
\hline
%
$0.01$ &SAT & $0.0035$& $0.0023$& $0.11$&$0.0046$ &$0.005$&$8.3\times  10^{-5}$&$0.00052$ &$0.0024$ &--& $0.0063$\\ 
$$ &FG10\% FLT & &&&&&&&&&\\ 
$0.01$ &+CR &$0.0013$ & $0.0023$& --&$0.0046$&$0.0049$&$8.3\times 10^{-5}$ &$0.0005$&$0.0023$& --&$0.0063$\\
\hline

$0.001$ &SAT FLT &$0.00051$ & $0.0023$& $0.13$&$0.0046$&$0.0050$&$8.3\times 10^{-5}$ &$0.00051$&$0.0023$& --&$0.0063$\\
$ $ &FG 10\% DASI50\% & &&&&&&&&&\\ 
\hline
$0.001$ &SAT FLT CR &$0.00042$ & $0.0023$& --&$0.0046$&$0.0050$&$8.3\times 10^{-5}$ &$0.00051$&$0.0023$&--&$0.0063$\\
$ $ &FG 10\% DASI50\% & &&&&&&&&&\\
 \hline
     &SAT NO NOISE & &&&&&&&&&\\
$0.0001$ &FG1\% FLT &$0.00007$&$0.0019$&$0.17$&$0.0037$&$0.005$&$7.4\times 10^{-5}$&$0.0022$&$0.0008$&--&$0.005$\\
       & DASI 50\%&$$&$$&$$&$$&$$&$$&$$&$$&&$$\\
$0.0001$ & +CR&$0.000057$&$0.0019$&--& $0.0037$&$0.005$&$7.4\times
 10^{-5}$&$0.0022$&$0.0008$&--&$0.005$\\
\hline \hline

\end{tabular}
\end{table*}

\begin{table*}
\begin{tabular}{cccccccccccc}
$r$& case & $\Delta r$ & $\Delta n$ & $\Delta n_t$& $\Delta dn/d\ln k$ &$\Delta Z$&$\Delta \omega_b$ &$\Delta \omega_c$ &$\Delta h$ &
  $\Delta \Omega_K$& $\Delta_A$ \\ \hline 
$0.01$ &QUIET & $0.09$& $0.029$& $4.2$&$0.047$ &$0.06$&$0.0007$&$0.005$
   &$0.026$ &--& $0.083$\\
$0.03$ & FG1\% &$0.31$ &$0.03$ &$4.$ &$0.048$ &$0.062$ &$0.0007$ &$0.005$ &$0.025$ &-- &$0.083$ \\
$0.1$ & FLT & $0.32$ & $0.03$ & $1.3$ & $0.48$ & $0.062$ & $0.0007$ & $0.005$ & $0.025$ & -- & $0.083$\\
\hline
$0.01$  &QUIET&$0.015$&$0.029$&--&$0.045$&$0.061$&$0.00071$&$0.005$&$0.024$&--&$0.08$\\
$0.03$ &FG1\% FLT&$0.019$&$0.03$&--&$0.045$&$0.062$&$0.0072$ &$0.005$&$0.025$&--&$0.08$ \\
$0.1$  & CR& $0.035$&$0.03$&--&$0.046$&$0.062$&$0.00072$&$0.005$&$0.024$&--&$0.08$\\
\hline
$0.01$&QUIET&$0.13$& $0.03$& $6.0$&$0.047$ &$0.063$&$0.0007$&$0.005$
   &$0.023$ &--& $0.08$\\
$0.03$ & FG 10\% & $0.32$ & $0.032$ & $5.1$ & $0.05$ & $0.063$ & $0.0007$ & $0.05$ & $0.025$ & -- & $0.08$\\
$0.1$  &FLT&$0.35$ & $0.032$&$1.4$&$ 0.05$&$0.063$&$0.0007$ &$0.005$ &$0.024$ &--&$0.08$ \\
\hline
$0.01$  &QUIET &$0.019$&$0.03$&--&$0.047$&$0.06$&$0.0007$&$0.005$&$0.023$&--&$0.08$\\
$0.03$ &FG10\%,FLT& $0.024$&$0.047$&--&$0.047$&$0.06$&$0.0007$&$0.005$&$0.023$&--&$0.08$ \\
$0.1$ &CR& $0.037$&$0.024$&--&$0.046$&$0.06$&$0.0007$&$0.005$&$0.023$&--&$0.08$\\
\hline \hline
$0.01$& PolarBeaR&$0.07$ &$0.018$&$3.7$&$0.037$&$0.049$&$0.0006$&$0.0043$&$0.018$&--&$0.06$\\
$0.03$& DASI50\%, Dust1\%&$0.1$&$0.018$&$2.1$&$0.037$&$0.051$&$0.00062$&$0.0044$&$0.019$&--& $0.06$\\
$0.1$ & FG1\% FLT&$0.1$  &$0.018$&$0.59$&$0.037$&$0.052$&$0.00062$&$0.0044$&$0.019$&--&$0.06$\\
\hline
$0.01$&PolarBeaR&$0.012$&$0.018$&--&$0.037$&$0.051$&$0.00061$&$0.0043$&$0.01$9&--&$0.06$\\
$0.03$&DASI50\%, dust1\%&$0.015$ & $0.018$&--&$0.037$&$0.051$&$0.00061$&$0.0043$&$0.019$&--&$0.06$\\
$0.1$  &FG1\% FLT CR &$0.023$&$0.018$&--&$0.037$&$0.052$&$0.00061$&$0.0043$&$0.019$&--&$0.06$\\

\hline \hline

$0.01$&QUIET+PolarBeaR&$0.036$&$0.017$&$2.0$&$0.033$&$0.058$&$0.00066$&$0.0046$&$0.02$&--&$0.067$\\
$0.03$&DASI50\%dust1\%& $0.08$&$0.02$&$1.7$&$0.039$&$0.063$&$0.00067$&$0.0049$&$0.021$&--&$0.067$\\
$0.1$ &FLT FG 1\%&$0.08$&$0.02$&$0.47$&$0.039$&$0.06$&$0.00064$&$0.0047$&$0.02$&--&$0.067$\\
\hline
$0.01$&QUIET+PolarBeaR&$0.006$&$0.015$&--&$0.03$&$0.058$&$0.00066$&$0.0046$&$0.02$&--&$0.067$\\
$0.03$&DASI50\%dust1\%&$0.009$&$0.02$&--&$0.039$&$0.063$&$0.00067$&$0.0048$&$0.021$&--&$0.067$\\
$0.1$ &FLT FG 1\% CR&$0.017$ &$0.02$&--&$0.039$&$0.06$&$0.00064$&$0.0046$&$0.02$&--&$0.067$\\
\hline\hline
      & QUIETBeaR & & & & & & & & & & \\
$0.01$  &DASI10\%, dust1\% &$0.017$ &$0.054$&$1.0$&$0.094$&$0.15$&$0.0016$&$0.01$&$0.05$&--&$0.19$\\
       & FG 1\% FLT & & & & & & & & & & \\
  $0.01$ & +CR&  $0.003$ & $0.054$&-- &$0.093$&$0.15$&$0.0016$&$0.01$&$0.05$&--&$0.19$\\
\hline
\hline
& SPIDER, $\tau=0.1$ & & & & & & & & & & \\
$0.01$&DASI50\%&$0.02$&$0.1$ &$0.44$ &$0.06$ &$0.007$ & $0.002$& $0.01$ &$0.04$ &--&$0.1$ \\
& FG 1\% FLT&& & & & & & & & & \\
$0.01$& +CR&$0.004$&$0.1$&-- &$0.06$&$0.007$&$0.002$&$0.009$&$0.03$&-- &$0.09$ \\
\hline
\end{tabular}
\end{table*}

\begin{table*}
 \caption{\label{table:experiments} Experimental specifications used
 for computing cosmological constraints.}
 \begin{tabular}{ccccc}
               &frequency & noise/beam & beam FWHM & sky coverage \\
               &  (GHz)   & ($\mu$K)   &  ($^\prime$)      & ($f_{sky}$ or sq. deg.)   \\
 \hline \hline 
SAT          &   30, 50, 70, 100, 200  &   2.2 per channel  &  8  &   $f_{sky}$=0.8  \\
\hline
QUIET          &   40  &   0.43  &  23  &   $4 \times 400$  \\
               &   90  &   0.78  &  10  &                   \\
\hline
PolarBeaR      &   90  &   1.6  &  6.7  &   500  \\
               &   150  &   2.4  &  4.0  &        \\
               &   220  &   11.3  &  2.7  &        \\
\hline
QUIET+PolarBeaR      &   40  &   0.43  &  23  &   $400$  \\
                     &   90  &   0.78  &  10  &          \\
               &   90  &   1.6  &  6.7  &     \\
               &   150  &   2.4  &  4.0  &        \\
               &   220  &   11.3  &  2.7  &        \\
\hline
QUIETBeaR      &   40  &   0.1  &  23  &   $170$  \\
                     &   90  &   0.18  &  10  &     \\
                     &   90  &   0.18  &  10  &     \\
                     &   150  &   0.18  &  10  &     \\
                     &   220  &   0.18  &  10  &     \\
\hline
SPIDER          &   40  &   0.74  &  145  &   $f_{sky}$=0.4  \\
                &   84  &   0.36  &  69  &   \\
                &   92  &   0.36  &  63  &   \\
                &   100  &   0.36  &  58  &   \\
                &   145  &   0.58  &  40  &   \\
                &   220  &   1.6  &  26  &   \\
\hline
\end{tabular}
\end{table*}

\end{document}